\documentclass[a4paper,11pt]{article}
\pdfoutput=1 

\usepackage{jcappub} 

\usepackage[T1]{fontenc} 
\usepackage{comment}

\title{\boldmath Gauge fixing in cosmological perturbations of Unimodular Gravity}

\author{Francisco X. Linares Cede\~no$^1$\note{Corresponding author.}}
\author{and Ulises Nucamendi}
\affiliation{Instituto de F\'isica y Matem\'aticas, Universidad Michoacana de San Nicol\'as de Hidalgo,\\ Edificio C--3, Ciudad Universitaria, CP. 58040 Morelia, Michoac\'an, M\'exico.}

\emailAdd{francisco.linares@umich.mx}
\emailAdd{ulises.nucamendi@umich.mx}

\abstract{With focus on the cosmological evolution of linear perturbations of matter and geometry, we calculate the equivalent expressions to that of the Newtonian and Synchronous gauges within the framework of Unimodular Gravity, being these two gauges commonly used and implemented in Boltzmann codes. An important aspect of our analysis is the inclusion of the energy--momentum current violation, as well as its perturbations. Moreover, for the first time we demonstrate that it is possible to fix both gauges consistently, although as it has been already noticed in previous literature, neither of them is recovered in the sense of the dynamics given in General Relativity for matter and metric fluctuations. Specifically, we show that since the unimodular constraint at the level of linear perturbations lead to only one degree of freedom of scalar modes of metric fluctuations, the dynamics in Unimodular Gravity forces to keep the anisotropic stress in the Newtonian gauge, whereas the cold dark matter comoving frame can not be set in the Synchronous gauge. The physical implications on the density contrast of cold dark matter is reviewed, and the Sachs--Wolfe effect is obtained and compared with previous results in the literature of cosmological perturbations in Unimodular Gravity.}

\begin{document}
\maketitle
\flushbottom

\section{Introduction}
\label{intro}

\qquad We are living in the era of modern cosmology, and accurate predictions of the dynamics of the universe are required. With the advent of more data, the information acquired from different sources in the cosmos challenges the broad variety of cosmological models in the literature. The standard \textit{Lambda Cold Dark Matter} ($\Lambda$CDM) model is based on the theory of \textit{General Relativity} (GR), which offers a great concordance with several observations~\cite{Turyshev:2008dr,Uzan:2010ri,Everitt:2011hp,Will:2018lln,LIGOScientific:2018dkp,EventHorizonTelescope:2020qrl,Krishnendu:2021fga}. However, there are still some unsolved problems within the $\Lambda $CDM model~\cite{Bull:2015stt,Bernal:2016gxb,DelPopolo:2016emo,Perivolaropoulos:2021jda,Sales:2022ich,DiValentino:2022fjm}. This has motivated a large amount of proposals considering both new particles and new theories of gravity.

The current accelerated expansion of the universe is one of the riddles that cosmologist are trying to solve, and within the framework of GR it is the \textit{Cosmological Constant} $\Lambda$ who plays the role of the main component responsible of such accelerated expansion, the so--called \textit{Dark Energy}. The origin of this dark component is still unknown, and there are many models in the literature based on different physics: some examples are dark energy fluids~\cite{Linder:2008ya,Bamba:2012cp,Luongo:2013eza,Bini:2014pmk,Chavanis:2015paa,Barrera-Hinojosa:2019yyh}, quintessence/phantom scalar fields~\cite{Elizalde:2004mq,Nojiri:2005sx,Dutta:2008px,Sheykhi:2011cn,Sola:2016hnq,Battye:2016alw,LinaresCedeno:2019bgo,Vazquez:2020ani,Urena-Lopez:2020npg,LinaresCedeno:2021aqk,Roy:2022fif,Vazquez:2023kyx}, modified gravity~\cite{Kunz:2006ca,Koyama:2006ef,Nojiri:2010wj,Capozziello:2010uv,Tsujikawa:2010zza,Battye:2013ida,Nojiri:2017ncd,Granda:2019agf,Jaber:2021hho,Wang:2022xdw}, between others.

In first formulations of GR, a choice of coordinates such that the determinant of the metric tensor is fixed was considered~\cite{Einstein:1914bx}, this is, the metric tensor $g_{\mu \nu}$ obeys the \textit{unimodular condition} $\sqrt{-g} = 1$.  Later, the relation between a fixed metric determinant and the cosmological constant was made~\cite{Anderson:1971pn}, where \textit{unimodular coordinate mappings} were imposed: $x^{\mu}\rightarrow x^{\prime \mu}$ such that $\mid \partial x^{\prime \mu}/\partial x^{\nu} \mid = 1$. Such consideration leads to the traceless part of the Einstein field equations, and this gravitational theory has been dubbed \textit{Unimodular Gravity} (UG)~\cite{Anderson:1971pn,vanderBij:1981ym,Weinberg:1988cp,Buchmuller:1988wx,Buchmuller:1988yn,Unruh:1988in,Henneaux:1989zc,Ng:1990xz}. One of the main consequences of having a four--volume preserving theory is that the energy--momentum tensor is no longer conserved ($\nabla_{\mu} T^{\mu}_{\ \nu} \neq 0$), and new non--gravitational interactions are allowed in the matter--energy sector. This feature is expected in theories of gravity that at a fundamental level could be more compatible with quantum mechanics~\cite{Maudlin:2019bje,Bonder:2022kdw}.

Cosmological models based on UG have been studied in last years~\cite{Alvarez:2007nn,Jain:2012gc,Ellis:2013uxa,Nojiri:2015sfd,Nojiri:2016ygo,Barvinsky:2017pmm,Garcia-Aspeitia:2019yni,Barvinsky:2019agh,Perez:2020cwa,Corral:2020lxt,LinaresCedeno:2020uxx,Singh:2023jxu}, although most of them have been focused in the background dynamics. Particularly in~\cite{LinaresCedeno:2020uxx}, the authors of the present work analyzed four phenomenological diffusion models describing interactions between the dark sector components, and it is shown that such interactions alleviate the $H_0$ tension. It is then natural to go further, and studying whether it is possible to describe an inhomogenous universe by considering linear perturbations in UG, with the aim of reproduce observables such as the \textit{Cosmic Microwave Background} (CMB) and the \textit{Large Scale Structure} (LSS). This implies to properly solve the Einstein--Boltzmann system describing the cosmological evolution for the initial fluctuations of both, the all matter components and the metric tensor.

Some contributions in this direction have been made: in~\cite{Gao:2014nia,Basak:2015swx} the linear perturbations are obtained although the authors impose by hand the conservation of the energy--momentum tensor, which is not the most general form of the UG field equations. Nonetheless, it is shown that the \textit{Sachs--Wolfe effect}~\cite{Sachs:1967er} in UG has a new term given by a scalar metric perturbation~\cite{Gao:2014nia}. On the other hand, the second order perturbations were obtained and no major distinctions from GR are found~\cite{Basak:2015swx}. A recent analysis including the non--conservation of the energy--momentum tensor has been realized in~\cite{Garcia-Aspeitia:2019yod,Fabris:2021atr,deCesare:2021wmk,Alvarenga:2023oep} and then, the presence of an \textit{energy--momentum current violation} has been considered.

As we mentioned above, whereas a first work was made by the authors of the present work with focus in the Hubble tension at a background level~\cite{LinaresCedeno:2020uxx}, now we will pay all the attention to the details of the cosmological perturbations in UG, which is a previous necessary step to the posterior implementation of this theory at the level of numerical solutions from Boltzmann solvers, as well as statistical analysis in order to both constraint parameters, and performing model comparison. A particular aspect that has not been mentioned in previous literature is that of gauge fixing. This is of crucial importance due to one have to ensure that no spurious degrees of freedom are propagating in the theory. We show that it is possible to fix both \textit{Newtonian} and \textit{Synchronous} gauges: the former is fixed in the same way as in GR, whereas for the latter to be fixed the unimodular constraint at first order is needed to be implemented. Nonetheless, the dynamics of the linear perturbations differs to that of GR due to both, the unimodular constraint at linear order, and the presence of the energy--momentum current violation. Thus, with the aim of exploring possible imprints of UG dynamics at first order, we obtain the CDM density contrast in both gauges mentioned above for a matter--dominated universe, and the Sachs--Wolfe effect is obtained as well.

This paper is organized as follows: Section~\ref{state_of_art} is dedicated to review in more detail the current contribution of previous works on the analysis of linear perturbations in UG. In Section~\ref{emug} we obtain the equations of motion for UG by variations method. Once the field equations are obtained, the background for a \textit{Friedmann--Robertson--Walker} (FRW) line element is implemented and the linear perturbations are obtained in Section~\ref{lin_pert_UG_eqs}. Besides, for the first time the perturbations for the energy--momentum current violation is presented. Later in Section~\ref{gauge_fixing_UG} we present the main analysis of this work, we fix the most commonly used gauges in cosmology: Newtonian and Synchronous gauges. Additionally, we review the gauge choice implemented at~\cite{Gao:2014nia,Basak:2015swx}. In Section~\ref{phys_imp} we present the physical implications of UG linear perturbations dynamics, and particularly we pay attention to the evolution of CDM density contrast, as well as in the Sachs--Wolfe effect. Final remarks are given in Section~\ref{final_remarks}.

We will follow the signature convention $(-\, , +\, , +\, , +)$. On the other hand, we will use dots (\ $\dot{}$\ ) for derivatives with respect to cosmic time $t$, and primes (\ $^{\prime}$\ ) will be used to label change of coordinates when choosing gauges.

\section{Unimodular Gravity perturbation theory: State of art}
\label{state_of_art}

\qquad In this Section we are going to briefly review previous works that have addressed the cosmological perturbations in UG. We will highlight the main results, as well as several crucial aspects that have not been deeply analyzed yet.

The first analysis on linear perturbations within the framework of UG was done by~\cite{Gao:2014nia}. In such work it is shown for the first time the unimodular constraint at linear perturbations level: what at the background level is a fixed four--volume, now is a new relation between the scalar modes of the metric fluctuations that is not present in GR at the level of linear perturbations. Notwithstanding, the unimodular constraint leads to a gauge issue that seems to be unavoidable when obtaining the dynamics for the scalar modes of the metric fluctuations. Specifically, the authors of~\cite{Gao:2014nia} explain:\\

\textit{"This scalar type metric perturbation cannot be removed through a gauge choice in unimodular gravity and thus leads to the possibility of observationally distinguishing unimodular gravity from GR"}.\\

In fact, such scalar mode appears as a new term when obtaining the relation between temperature and gravitational potential in the Sachs--Wolfe effect within the framework of UG. However, the differences between GR and UG when analyzing the Sachs--Wolfe effect are suppressed on large angular scales, making both theories practically indistinguishable. It is important to mention that even when the unimodular constraint gives an additional relation between the gravitational potentials for the metric perturbations, the gauge choice made by the authors of~\cite{Gao:2014nia} leaves two gravitational degrees of freedom, as in GR. This is done so in order to compare with the longitudinal gauge of GR. However, this choice does not fix the gauge, as it will be shown in the present work.

Later, going a step further the authors of~\cite{Basak:2015swx} obtain the second order perturbations of the theory. The gauge choice they consider is the same as that in~\cite{Gao:2014nia}, and then the gauge fixing issue persists.  Nonetheless, the authors of~\cite{Basak:2015swx} claim that the appearance of the new term in the Sachs--Wolfe effect can be compatible with GR with the proper gauge choice. On the other hand, the second order Mukhanov--Sasaki equation in UG is obtained, and it is shown that it depends only on the first order unimodular constraint. It is concluded that there is no significant difference between GR and UG at neither of both first and second order in perturbations.

Then, both works mentioned above conclude that there are no major distinctions between GR and UG. However, in both studies it is assumed that the energy--momentum tensor is conserved, which neglect one of the main features of the UG theory. Current progress in this direction has been done: in~\cite{Garcia-Aspeitia:2019yod} the cosmological linear perturbations in UG under the Newtonian conformal gauge are studied for scalar and tensor perturbations, and the Boltzmann equation for photons is obtained as well. In particular, they obtain the $00$ component for scalar perturbations, an it is shown that it presents an extra contribution due to $\Lambda$. On the other hand, the Boltzmann equation for photons contains an additional term that contains third order derivatives. In this respect, the authors say:\\

"\textit{...the Boltzmann equation for photons is exposed because it contains the energy momentum violations that characterize the UG. Notoriously, the extra term carries higher order derivatives in the conformal time component for the scale factor and for the scalar curvature.}"\\

This is the case when considering that radiation will be coupled with non--standard terms due to the energy--momentum current violation. In our case, we will consider that the new non--gravitational interactions occur only between the dark sector components. This is so in order to keep the same philosophy as in our previous work~\cite{LinaresCedeno:2020uxx}, where as a conservative assumption we considered that the interactions of the ordinary matter are ruled by the laws of the standard model of particle physics, which are very well known and tested~\cite{ParticleDataGroup:2022pth} at least in comparison with the knowledge we have about the possible properties of both dark matter and dark energy.\footnote{While to date there is no detection of dark matter/dark energy in terrestrial laboratories and then no signal of ordinary matter/dark component interactions has been registered, it is possible to find in the literature several proposals of possible interactions between radiation and baryons with the components of the dark sector through astrophysical and cosmological sources within the framework of GR~\cite{Dvorkin:2013cea,Tashiro:2014tsa,Schewtschenko:2014fca,Munoz:2015bca,Barkana:2018lgd,Vagnozzi:2019kvw,Ferlito:2022mok}.}

Another work assuming violation of the energy--momentum tensor as well as considering linear perturbations in the longitudinal gauge in UG is~\cite{deCesare:2021wmk}, where the authors do not use the action principle to obtain the UG field equations, but their starting point is the trace-free Einstein equations only. Thus, the authors do not assume the unimodular condition. This is different from our approach where the UG field equations are derived from an action principle (in fact, we will consider unimodular variations), and the unimodular constraint will be considered for both, background and linear perturbations. The study of instabilities in UG with non--gravitational interactions in the dark sector are carried out with detail, and in particular the authors report:\\

"\textit{...the usual instability is driven by the nonadiabatic pressure perturbation of the dark energy fluid, but for the trace--free Einstein equations and a transfer potential that depends only on the dark matter energy density there is no nonadiabatic pressure perturbation to dark energy — this is ultimately why there is no instability here.}"\\

Later in~\cite{Fabris:2021atr,Alvarenga:2023oep} the contribution of an energy--momentum current violation is considered as well, and it is referenced by the authors as \textit{nonconservative unimodular gravity}. Besides, the unimodular constraint is fully considered when obtaining the dynamics of linear perturbations, which are written in terms of one only gravitational degree of freedom. In relation to the Newtonian and Synchronous gauges, the authors in~\cite{Fabris:2021atr} report:\\

"\textit{The newtonian gauge can not be used in the unimodular  context  unless  any  anisotropic  contribution  to the stress--tensor are considered,}" and \\

"\textit{Scalar perturbations in the nonconservative unimodular gravity are permitted in the synchronous gauge only and have a growing mode.}"\\

The former aspect is still without analysis, whereas the latter was analyzed by considering a specific background solution. Even when these results constitute a significant breakthrough in the study of cosmological perturbations in UG, there is still missing the proper analysis of gauge fixing, and looking for the choices that leave the theory without spurious degrees of freedom. Moreover, it is mandatory to have the correct dynamics of linear perturbations if one is interested in the implementation of UG cosmological model in Boltzmann solvers such as \texttt{CLASS}~\cite{Lesgourgues:2011re} and \texttt{CAMB}~\cite{Lewis:1999bs}. For instance, \texttt{CAMB} integrates
transfer functions for quantities
defined in the synchronous gauge, whereas \texttt{CLASS} uses the synchronous gauge by default although Newtonian gauge equations are implemented on top of the synchronous ones. Therefore, the gauge fixing issue in UG must be deeply understood\footnote{Since Boltzmann solvers mentioned above are still being used by cosmologist, we are interested in the study of Newtonian/Synchronous gauges. The gauge invariant formalism of UG is presented in detail in~\cite{Gao:2014nia,Basak:2015swx,deCesare:2021wmk,Alvarenga:2023oep}, and it is not our focus to deal which such treatment in the present work.}. In this regard, it has been recently reported in~\cite{Landau:2022mhm} the implementation of a diffusion model in UG to obtain the CMB anisotropies by using \texttt{CAMB}. The model consists in postulating an effective cosmological constant energy density, with which is possible to obtain a modified matter energy density for both baryons and dark matter. However, the authors use the same set of equations of GR at linear level of perturbations, and they argue:\\

"\textit{We did not consider the linear perturbation equations of unimodular gravity, because we assume that the corresponding modification in the observable quantities would be very tiny compared to the change introduced by the background quantities.}"\\

Even when this is a reasonable assumption from the modeling point of view, the technical aspects involving the use of \texttt{CAMB} demand a meticulous analysis of the gauge in which the linear perturbations are written. As we said before, \texttt{CAMB} calculates cosmological perturbations in the synchronous gauge, and as we will show in detail in this work, such gauge can not be imposed within the framework of UG at least in the same way as it is done in GR. In this sense, the analysis of~\cite{Landau:2022mhm} can be thought of as having a theory of gravity for the background dynamics (UG), and another one for linear perturbations (GR). Thus, despite the fact that this represents a first approach to reproduce CMB anisotropies in UG, the numerical solutions and posterior distributions obtained by the authors of such work lead to results whose physical interpretation has to be taken carefully.

In a more recent work, it is claimed that it is a common mistake to impose the condition of a fixed four--volume at the level of the perturbations~\cite{Bengochea:2023dep}. Moreover, it is said that\\

"\textit{if we were to follow the standard literature on the subject, e.g. Refs. [7, 9] (~\cite{Gao:2014nia,Basak:2015swx} in the present manuscript), and erroneously impose the condition mistakenly described as the "UG constraint", namely, $g^{\mu \nu }\delta g_{\mu \nu } = 0$, we would have been prevented from working with the Newtonian gauge.}"\\

As we will show below, it is possible to work with the Newtonian gauge when the UG constraint the authors mention is considered. In fact, we will show that such gauge is fixed in the same way as GR, but the dynamics of perturbations will have new features in the matter sector.

From all mentioned above, there are some important technical details still to be addressed in order to properly study the cosmological perturbations in UG, with aim of analyze whether cosmological models based on UG are viable candidates to describe the universe, not only the background dynamics but to describe CMB and LSS as well.

Summarizing, we have that
\begin{itemize}
    \item Even when UG naturally leads to non--gravitational interactions, this is $\nabla_{\mu} T^{\mu}_{\ \ \nu} \neq 0$, first works on UG linear perturbations assume the opposite, and energy--momentum conservation is imposed by hand~\cite{Gao:2014nia,Basak:2015swx}. Within these analysis, GR and UG are basically the same theory.
    \item With focus on scalar modes, once the non--conservation of energy--momentum tensor is considered it has been possible to obtain the $00$ component of field equations for linear perturbations in the Newtonian gauge~\cite{Garcia-Aspeitia:2019yod}, the perturbed field equations in the longitudinal gauge but without considering the unimodular constraint~\cite{deCesare:2021wmk}, and solutions for linear perturbations only in the synchronous gauge~\cite{Fabris:2021atr,Alvarenga:2023oep}. Newtonian gauge requires anisotropic term in the matter sector, and solutions in this gauge are lacking.
    \item The progress reported at~\cite{Landau:2022mhm,Bengochea:2023dep} is done under the claim that it is not necessary to hold the condition of having a fixed four--volume at the level of perturbations. In our case, we will consider that such geometric condition has to be implemented as is the case for the violation of the energy--momentum tensor conservation, in the sense that both features (the former geometric and the latter in the matter sector) are intrinsic ingredients of the UG theory.
    \item None of the works mentioned above (except ~\cite{Bengochea:2023dep}) analyze the gauge fixing in UG. This is crucial to do in order to avoid the propagation of spurious degrees of freedom that can be mistaken with physical effects on cosmological observables.
\end{itemize}

In the present work we will address the missing points mentioned above, with special emphasis on the gauge fixing problem, the dynamics of the evolution of linear perturbations considering the unimodular constraint as well as the energy--momentum current violation, and the physical repercussions on the growth of CDM density contrast. In this respect, we show that it is possible to fix both the Newtonian and Synchronous gauges, although they have different consequences in the dynamics of linear perturbations. We also review the gauge choice studied at~\cite{Gao:2014nia,Basak:2015swx}, and we show that it is not completely fixed due to a remaining undetermined function. We find analytical solutions in terms of the only gravitational degree of freedom in Newtonian and Synchronous gauges. Therefore, a possibility to track some signatures of UG at cosmological scales is by analyzing the growth of structure, which implies to obtain the dynamics not only for CDM, but for all the matter components and by setting the proper dynamical equations for the Einstein--Boltzmann system within the framework of UG. On the other hand, we obtain the Sachs--Wolfe effect in UG, and different from what has been previously reported in the literature, the result is exactly the same as GR without new contributions of metric perturbations.

\section{Equations of motion in Unimodular Gravity}
\label{emug}

\qquad Different from~\cite{LinaresCedeno:2020uxx}, where the unimodular constraint was introduced in the Einstein--Hilbert action through a Lagrange multiplier, this time we will obtain the UG equations of motion considering the following \textit{unimodular variation} $\delta_u$,
\begin{equation}
    \delta_{u}g^{\mu \nu} \equiv \delta g^{\mu \nu} - \frac{1}{4}g^{\mu \nu}g_{\alpha \beta} \delta g^{\alpha \beta}\, ,
    \label{uni_var}
\end{equation}
and then, it follows that
\begin{equation}
    g_{\mu \nu}\delta_{u}g^{\mu \nu} = g_{\mu \nu}\left( \delta g^{\mu \nu} - \frac{1}{4}g^{\mu \nu}g_{\alpha \beta} \delta g^{\alpha \beta} \right) = g_{\mu \nu} \delta g^{\mu \nu} - \frac{1}{4}\delta_{\lambda}^{\lambda}g_{\mu \nu} \delta g^{\mu \nu}= 0\, .
\end{equation}

The volume--preserving diffeomorphisms are satisfied under the unimodular variation $\delta_u$, since when considering the variation of the determinant of the metric, we have
\begin{equation}
    \delta_u\sqrt{-g} = -\frac{1}{2}\sqrt{-g}g_{\mu \nu}\delta_u g^{\mu \nu} = 0\quad\Rightarrow \quad \sqrt{-g} = f\, ,
    \label{vol_diff}
\end{equation}
where $f=f(x)$ is a nondynamical scalar density which depends on the coordinates, and it can always be redefined to the unity.

The total action is,
\begin{equation}
    S = S_{EH} + S_M = \frac{1}{2\kappa^2}\int d^4x\sqrt{-g}R + S_M\, ,
    \label{full_action}
\end{equation}
where $S_{EH}$ is the Einstein--Hilbert action, and $S_M$ is the action for the matter fields. The Ricci scalar is defined as $R=g^{\mu \nu}R_{\mu \nu}$, and then, the unimodular variation of the Einstein--Hilbert action is
\begin{equation}
    \delta_u S_{EH} = \frac{1}{2\kappa^2}\left[\int d^4x\left(\delta_u \sqrt{-g}\right)R + \int d^4x \sqrt{-g}\left(\delta_u g^{\mu \nu}\right)R_{\mu \nu} + \int d^4x \sqrt{-g}g^{\mu \nu}\left(\delta_u R_{\mu \nu}\right)\right]\, .
\end{equation}

The first term is zero due to~\eqref{vol_diff}, and the last term is also vanishing because, as in GR, after some algebra such term is a boundary contribution at infinity which can be set to zero~\cite{Carroll:2004st}. Then, the Einstein--Hilbert action gets reduced to the second term only, which using Eq.~\eqref{uni_var} is written as
\begin{eqnarray}
    \delta_u S_{EH} &=& \frac{1}{2\kappa^2}\int d^4x \sqrt{-g}\left(\delta_u g^{\mu \nu}\right)R_{\mu \nu} = \frac{1}{2\kappa^2}\int d^4x \sqrt{-g}\left(\delta g^{\mu \nu} - \frac{1}{4}g^{\mu \nu}g_{\alpha \beta} \delta g^{\alpha \beta}\right)R_{\mu \nu}\, , \nonumber \\
    &=& \int d^4x \sqrt{-g}\left[\frac{1}{2\kappa^2}\left(R_{\mu \nu} - \frac{1}{4}R g_{\mu \nu} \right)\right]\delta g^{\mu \nu}\, .
    \label{eh_action}
\end{eqnarray}

For the matter content, we have the standard energy--momentum tensor definition but considering the unimodular variation, this is,
\begin{equation}
    T_{\mu \nu}\equiv -\frac{2}{\sqrt{-g}}\frac{\delta_u S_{M}}{\delta_u g^{\mu \nu}}
\end{equation}
and then, we have
\begin{eqnarray}
     \delta_u S_{M}&=& -\frac{1}{2}\sqrt{-g}T_{\mu \nu}\delta_u g^{\mu \nu} = -\frac{1}{2}\sqrt{-g}T_{\mu \nu}\left(\delta g^{\mu \nu} - \frac{1}{4}g^{\mu \nu}g_{\alpha \beta} \delta g^{\alpha \beta}\right) \nonumber \\
     &=& -\frac{1}{2}\sqrt{-g}\left( T_{\mu \nu} - \frac{1}{4}T g_{\mu \nu}  \right)\delta g^{\mu \nu}\, .
     \label{m_action}
\end{eqnarray}

Therefore, the previous results from Eq.~\eqref{eh_action} and~\eqref{m_action} give the following variation for the total action~\eqref{full_action},
\begin{equation}
    \delta_u S = \int d^4x \frac{\sqrt{-g}}{2}\left[\frac{1}{\kappa^2}\left(R_{\mu \nu} - \frac{1}{4}R g_{\mu \nu} \right) -\left( T_{\mu \nu} - \frac{1}{4}T g_{\mu \nu}  \right)\right]\delta g^{\mu \nu} = 0\, ,
\end{equation}
and thus, we obtain the UG field equations,
\begin{equation}
    R_{\mu \nu} - \frac{1}{4}R g_{\mu \nu} = \kappa^2\left( T_{\mu \nu} - \frac{1}{4}T g_{\mu \nu}  \right)\, ,
    \label{ug_eq_1}
\end{equation}
which are the trace--free version of the Einstein field equations. We can rewrite Eq.~\eqref{ug_eq_1} as follows,
\begin{equation}
    R^{\mu}_{\ \ \nu} - \frac{1}{2}R\delta^{\mu}_{\nu} + \frac{1}{4}\left(R + \kappa^2T\right)\delta^{\mu}_{\nu} = \kappa^2 T^{\mu}_{\ \ \nu}\, ,
    \label{ug_eq_efe}
\end{equation}
and applying the Bianchi identities,
\begin{equation}
    \nabla_{\mu}\left( R^{\mu}_{\ \ \nu} - \frac{1}{2}R\delta^{\mu}_{\ \ \nu} \right) + \frac{1}{4}\nabla_{\nu}\left(R + \kappa^2T\right) = \kappa^2 \nabla_{\mu}T^{\mu}_{\ \ \nu}\, ,
\end{equation}
we notice that whereas the first term on the l.h.s. is identically zero, the covariant derivative of the energy--momentum tensor is no longer locally conserved,
\begin{equation}\label{eff_cc}
    \kappa^2\nabla_{\mu}T^{\mu}_{\ \ \nu} = \frac{1}{4}\partial_{\nu}\left(R + \kappa^2T\right) \equiv J_{\nu}\, ,
\end{equation}
where $J_{\nu}$ is the \textit{energy--momentum current violation}. Due to this result, we will highlight in our discussion the presence of $J_{\nu}$ in the equations obtained from the covariant derivative of the energy--momentum tensor by using the expression \textit{(non) conservation}, which is a main distinctive feature of UG when compared with GR. Thus, the energy--momentum tensor is conserved when $J_{\nu} = 0$, and it is not conserved when $J_{\nu} \neq 0$. Along this work, we will solve the dynamics of UG considering a non--null energy--momentum current violation, and it will be only to compare with GR that $J_{\nu} = 0$ will be used. Integrating the expression from above, and replacing this result into Eq.~\eqref{ug_eq_efe}, we have
\begin{equation}
    R_{\mu \nu} - \frac{1}{2}Rg_{\mu \nu} + \left[ \Lambda + \int dx^{\alpha} J_{\alpha}(x) \right] g_{\mu \nu} = \kappa^2 T_{\mu \nu}\, ,
    \label{EFE_UG_gen}
\end{equation}
where $\Lambda$ is a constant of integration. Notice that, in the particular case when the energy--momentum tensor is conserved ($J_{\nu} = 0$), Eq.~\eqref{EFE_UG_gen} coincides with the Einstein field equations of GR, and then, $\Lambda$ is identified as the cosmological constant. Thus, within the framework of UG, the cosmological constant $\Lambda$ arises naturally in the equation of motion as an integration constant when considering volume--preserving diffeomorphisms. Notwithstanding, in general we will have $J_{\nu} \neq 0$, and non--gravitational interactions are allowed between different matter and energy components.

In summary, the UG field equations are given by
\begin{subequations}\label{ug_eqs}
\begin{eqnarray}
     R_{\mu \nu} - \frac{1}{2}Rg_{\mu \nu} + \Lambda(x) g_{\mu \nu} &=& \kappa^2 T_{\mu \nu}\, ,\quad {\rm{with}}\quad \Lambda(x) \equiv \Lambda + \int dx^{\alpha} J_{\alpha}(x)\, , \label{ug_eq_eff_cc}\\
     \nabla_{\mu} T^{\mu}_{\ \ \nu} &=& \frac{1}{\kappa^2}J_{\nu}\, ,\quad {\rm{with}}\quad J_{\nu}\equiv \frac{1}{4}\partial_{\nu}\left(R + \kappa^2T\right)\, ,\label{em_vio}
\end{eqnarray}
\end{subequations}
where $\Lambda(x)$ in Eq.~\eqref{ug_eq_eff_cc} is an \textit{effective cosmological constant} which in general depends on the spacetime coordinates. As we have mentioned in Section~\ref{state_of_art}, we will focus in non--gravitational interactions only between dark matter and the effective cosmological constant through the energy--momentum current violation $J_{\nu}$ according to Eq.~\eqref{em_vio}.

\section{Linear cosmological perturbations in Unimodular Gravity}\label{lin_pert_UG_eqs}

\qquad Let us write the metric, the energy--momentum tensor, the effective cosmological constant, and the energy--momentum current violation in the following way
\begin{equation}
g_{\mu \nu} = \bar{g}_{\mu \nu} + h_{\mu \nu}\, , \quad T_{\mu \nu} = \bar{T}_{\mu \nu} + \delta T_{\mu \nu}\, , \quad \Lambda = \bar{\Lambda} + \delta \Lambda\, ,\quad J_{\mu} = \bar{J}_{\mu} + \delta J_{\mu}\, ,
\label{pert_quant}
\end{equation}
where the bar denotes quantities from the background, and $h_{\mu \nu}\, , \delta T_{\mu \nu}\, , \delta \Lambda\, ,$ and $\delta J_{\mu}$ are small fluctuations with respect to their corresponding background values. In the case of the background metric, we consider the flat FRW spacetime, whose components are
\begin{subequations}\label{flrw}
\begin{eqnarray}
    \bar{g}_{00} &=& -1\, ,\quad \bar{g}_{0i} = 0\, ,\quad \bar{g}_{ij} = a^2(t)\delta_{ij}\, ,\\
    \bar{g}^{00} &=& -1\, ,\quad \bar{g}^{0i} = 0\, ,\quad \bar{g}^{ij} = a^{-2}(t)\delta_{ij}\, ,
\end{eqnarray}
\end{subequations}
while for the inverse of the metric perturbation we have
\begin{equation}
    h^{\mu \nu} = -\bar{g}^{\mu \alpha}\bar{g}^{\nu \beta}h_{\alpha \beta}\, ,
\end{equation}
whose components are given by
\begin{equation}
    h^{00} = -h_{00}\, ,\quad h^{i0}=a^{-2}h_{i0}\, ,\quad h^{ij}=-a^{-4}h_{ij}
\end{equation}

Notice that the determinant of the metric given by Eq.~\eqref{pert_quant} can be written at first order as
\begin{equation}
    \sqrt{-g}\simeq \sqrt{-\bar{g}}\left[ 1 + \frac{1}{2}\bar{g}^{\mu \nu}h_{\mu \nu} + \mathcal{O}(h^2) \right] = \sqrt{-\bar{g}}\left( 1 - \frac{h_{00}}{2} + a^{-2}\frac{h_{ii}}{2} \right)\, ,
\end{equation}
and, whereas at zero order we recover the unimodular constraint~\eqref{vol_diff}, at first order we have
\begin{equation}
    -h_{00} + a^{-2}h_{ii} = 0\, .
    \label{vol_diff_pert}
\end{equation}

The last expression will be important to be considered in the following analysis of the dynamics of small fluctuations, since it constitutes a new relation between the components of the perturbed metric that is not present in GR.

The Christoffel symbols are defined as
\begin{equation}
    \Gamma^{\alpha}_{\mu \nu} = \frac{1}{2}g^{\alpha \beta}\left( \partial_{\nu}g_{\beta \mu} + \partial_{\mu}g_{\beta \nu} - \partial_{\beta}g_{\mu \nu} \right)\, ,
\end{equation}
and then, the non--null components are
\begin{subequations}
\begin{eqnarray}
     \Gamma^0_{00} &=& -\frac{\dot{h}_{00}}{2}\, ,\\
     \Gamma^0_{i0} &=& \frac{\dot{a}}{a}h_{i0}-\frac{1}{2}\partial_i h_{00}\, ,\\
     \Gamma^0_{ij} &=& a\dot{a}\delta_{ij} + \frac{1}{2}\left( 2a\dot{a}\delta_{ij}h_{00} - \partial_j h_{i0} - \partial_i h_{j0} + \dot{h}_{ij} \right)\, ,\\
     \Gamma^i_{00} &=& \frac{1}{2a^2}\left( 2\dot{h}_{i0} - \partial_i h_{00} \right)\, ,\\
     \Gamma^i_{j0} &=& \frac{\dot{a}}{a}\delta_{ij} + \frac{1}{2a^2}\left( -2\frac{\dot{a}}{a}h_{ij} + \dot{h}_{ij} + \partial_j h_{i0} - \partial_i h_{j0} \right)\, , \\
     \Gamma^i_{jk} &=& \frac{1}{2a^2}\left( -2a\dot{a}h_{i0}\delta_{jk} + \partial_k h_{ij} + \partial_j h_{ik} - \partial_i h_{jk} \right)\, .
\end{eqnarray}
\end{subequations}

The Ricci tensor is,
\begin{equation}
    R_{\mu \nu} = \partial_{\alpha}\Gamma^{\alpha}_{\mu \nu} - \partial_{\nu}\Gamma^{\alpha}_{\alpha \mu} + \Gamma^{\alpha}_{\alpha \beta} \Gamma^{\beta}_{\mu \nu} - \Gamma^{\alpha}_{\nu \beta} \Gamma^{\beta}_{\alpha \mu}\, ,
\end{equation}
with components given by
\begin{subequations}
\begin{eqnarray}
     R_{00} &=& -3\frac{\ddot{a}}{a} - \frac{\nabla^2h_{00}}{2a^2} - \frac{3}{2}\frac{\dot{a}}{a}\dot{h}_{00} + \frac{\partial_i\dot{h}_{i0}}{a^2} -
     \frac{1}{2a^2}\left\lbrace \ddot{h}_{ii} - 2\frac{\dot{a}}{a}\dot{h}_{ii} + 2\left[ \left( \frac{\dot{a}}{a}\right)^2 - \frac{\ddot{a}}{a} \right]h_{ii} \right\rbrace\, ,\nonumber  \\
     \\
     R_{0i} &=& -\frac{\dot{a}}{a}\partial_i h_{00} - \frac{1}{2a^2}\left( \nabla^2 h_{i0} - \partial_i \partial_j h_{j0}\right) + \left[ \frac{\ddot{a}}{a} + 2\left( \frac{\dot{a}}{a} \right)^2 \right]h_{i0} - \frac{1}{2}\partial_t\left[ \frac{1}{a^2}\left( \partial_i h_{jj} - \partial_j h_{ji} \right) \right]\, ,\nonumber  \\
     \\
     R_{ij} &=& \left( a\ddot{a} + 2\dot{a}^2 \right)\delta_{ij} + \frac{1}{2}\partial_i\partial_j h_{00} + \left( 2\dot{a}^2 + a\ddot{a} \right)\delta_{ij} h_{00} + \frac{1}{2}a\dot{a}\delta_{ij}\dot{h}_{00} + \frac{1}{2}\ddot{h}_{ij} - \frac{\dot{a}}{a}\delta_{ij}\partial_k h_{k0}\nonumber \\
     && -\frac{1}{2a^2}\left( \nabla^2h_{ij} - \partial_k\partial_j h_{ki} - \partial_k\partial_i h_{kj} + \partial_i\partial_j h_{kk} \right) - \frac{1}{2}\frac{\dot{a}}{a}\left( \dot{h}_{ij} - \delta_{ij}\dot{h}_{kk} \right)\nonumber \\
     && +\left( \frac{\dot{a}}{a} \right)^2\left( 2h_{ij} - \delta_{ij}h_{kk} \right) - \frac{1}{2}\left( \partial_i \dot{h}_{j0} + \partial_j \dot{h}_{i0} \right) - \frac{1}{2}\frac{\dot{a}}{a}\left( \partial_i h_{j0} + \partial_j h_{i0} \right)\, ,
\end{eqnarray}
\end{subequations}

The Ricci scalar $R = \bar{R} + \delta R = \bar{g}^{\mu \alpha}\bar{R}_{\alpha \mu} + \bar{g}^{\mu \alpha}\delta R_{\alpha \mu} + h^{\mu \alpha}\bar{R}_{\alpha \mu}\, ,$ is given by
\begin{eqnarray}\label{ricci_sc}
     R &=& 6\left[ \left(\frac{\dot{a}}{a}\right)^2 + \frac{\ddot{a}}{a} \right] + 6\left[\left(\frac{\dot{a}}{a}\right)^2 + \frac{\ddot{a}}{a} \right]h_{00} + 3\frac{\dot{a}}{a}\dot{h}_{00} + \frac{\nabla^2h_{00}}{a^2} \nonumber \\
     && -\frac{2}{a^2}\left( 2\frac{\dot{a}}{a}\partial_ih_{i0} + \partial_i\dot{h}_{i0} \right) -\frac{2}{a^2}\left[ \left(\frac{\dot{a}}{a}\right)^2 + \frac{\ddot{a}}{a} \right]h_{ii} + \frac{\ddot{h}_{ii}}{a^2} \nonumber \\
     && -\frac{1}{a^4}\left( \nabla^2h_{ii} - \partial_i\partial_jh_{ij} \right)\, .
\end{eqnarray}

For the matter content we are going to be interested in the energy--momentum tensor of a perfect fluid, this is
\begin{equation}
    T_{\mu \nu} = \left( \rho + p \right)U_{\mu}U_{\nu} + pg_{\mu \nu}\, ,
    \label{emt}
\end{equation}
with
\begin{equation}
    \rho = \Bar{\rho} + \delta \rho\, ,\quad p = \Bar{p} + \delta p\, ,\quad U^{\mu} = \left( 1+\delta U^0, v^i \right)\, ,\quad U_{\mu} = (-1+\delta U_0,v_i)\, .
    \label{rpu_co}
\end{equation}
where $\rho$ is the energy density, $p$ the pressure, and $U^{\mu}$ the four--velocity of the fluid which at the level of the background we have chosen the system of reference of comoving observers. The term  $v^i=\delta U^{i}$ is the peculiar velocity, which can be considered as a small quantity as $\delta \rho$ and $\delta p$. Notice that, due to the condition $g_{\mu \nu}U^{\mu} U^{\nu}=-1$, the time component of the four--velocity perturbation is $\delta U^{0}=\delta U_0 = h_{00}/2$. Thus, the components of the energy--momentum tensor in terms of the zero and first order perturbations for a perfect fluid are given by
\begin{subequations}\label{tpert}
\begin{eqnarray}
T_{00} &=& \bar{\rho} -\bar{\rho} h_{00} + \delta \rho\, , \label{t00pert} \\
T_{i0} &=& \bar{p} h_{i0} - (\bar{\rho} + \bar{p})v_i\, , \label{t0ipert} \\
T_{ij} &=& a^2\bar{p}\delta_{ij} + \bar{p}h_{ij} + a^2\delta p\delta_{ij}\, , \label{tijpert}
\end{eqnarray}
\end{subequations}
whereas the components with mixed indices are
\begin{subequations}\label{mix_ind_emt}
\begin{eqnarray}
     T^0_{\ \ 0} &=& -\bar{\rho} - \delta \rho\, , \\
     T^0_{\ \ i} &=& -(\bar{\rho} + \bar{p})v_i = -T^{i}_{\ \ 0}\, ,\\
     T^i_{\ \ j} &=& \bar{p}\delta^i_j + \delta p \delta^i_j\, .
\end{eqnarray}
\end{subequations}

Once inserted the above expressions in the UG field equations \eqref{ug_eqs} for a spatially flat FRW universe, we obtain for the zeroth--order perturbations the background equations, i.e.,
\begin{subequations}\label{bg_eqs}
\begin{eqnarray}
     H^2 - \frac{\bar{\Lambda}(t)}{3} = \frac{\kappa^2}{3}\bar{\rho}\, &,&\label{friedmann}\quad 
     \dot{H} = -\frac{\kappa^2}{2}\bar{\rho}\left( 1 + \omega \right)\, ,\\
     \dot{\bar{\rho}} + 3H\bar{\rho}( 1 &+& \omega) = -\frac{\bar{J}_0(t)}{\kappa^2}\, ,\label{em_cons_bg}
\end{eqnarray}
\end{subequations}
where $H=\dot{a}/a$ is the Hubble parameter. Both $\bar{\Lambda}$ and $\bar{J}_0$ depend only on the cosmic time $t$ due to homogeneity and isotropy. The energy density and the pressure for the matter fields are related by a constant equation of state $\omega \equiv \bar{p}/\bar{\rho}\ (\omega=0$ for non--relativistic matter such as baryons and cold dark matter, and $\omega=1/3$ for radiation).

On the other hand, following~\cite{Weinberg:2008zzc} the linear perturbations for Eq.~\eqref{ug_eq_eff_cc} are
\begin{equation}
    \delta R_{\mu \nu} - \bar{\Lambda}h_{\mu \nu} - \bar{g}_{\mu \nu}\delta \Lambda = \kappa^2\left( \delta T_{\mu \nu} - \frac{1}{2}\bar{g}_{\mu \nu}\delta T - \frac{1}{2}h_{\mu \nu}\bar{T} \right)\, ,
    \label{1or_pert_ug}
\end{equation}
where $\bar{T}$ is the trace of the background energy--momentum tensor, and $\delta T$ its perturbation,
\begin{equation}
    \bar{T} = 3\bar{p} - \bar{\rho} = -\frac{6}{\kappa^2}\left[ \frac{\ddot{a}}{a} + \left(\frac{\dot{a}}{a}\right)^2 -\frac{2}{3}\bar{\Lambda} \right]\, ,\quad \delta T = 3\delta p - \delta \rho\, .
    \label{trace}
\end{equation}

The components of Eq.~\eqref{1or_pert_ug} are given by
\begin{subequations}
\begin{eqnarray}
\frac{\kappa^2}{2}\left( \delta \rho + 3\delta p \right) &=& -\frac{\nabla^2h_{00}}{2a^2} - \frac{3}{2}H\dot{h}_{00} + \frac{\partial_i\dot{h}_{i0}}{a^2} - 3\left( H^2 + \dot{H} \right)h_{00}\nonumber \\
&&- \frac{1}{2a^2}\left( \ddot{h}_{ii} - 2H\dot{h}_{ii} - 2 \dot{H} h_{ii} \right) + \delta\Lambda\, , \label{g00pert}  \\
-\kappa^2\left( \bar{\rho} + \bar{p} \right)v_i &=& -H\partial_i h_{00} - \frac{1}{2a^2}(\nabla^2h_{i0} - \partial_i\partial_jh_{j0}) - \frac{1}{2}\frac{\partial}{\partial t}\left[ \frac{1}{a^2}(\partial_i h_{jj} - \partial_j h_{ji}) \right]  \, ,\nonumber\\ \label{gi0pert} \\
\frac{a^2}{2}\kappa^2\left( \delta \rho - \delta p \right)\delta_{ij} &=& \frac{1}{2}\partial_i\partial_j h_{00} + (2\dot{a}^2+a\ddot{a})\delta_{ij}h_{00} + \frac{1}{2}a\dot{a}\delta_{ij}\dot{h}_{00} - \frac{H}{2}(\partial_ih_{j0} + \partial_jh_{i0}) \nonumber \\
&&- \frac{1}{2a^2}( \nabla^2h_{ij} - \partial_k\partial_j h_{ki} - \partial_k\partial_i h_{kj} + \partial_i\partial_jh_{kk} ) + \frac{1}{2}\ddot{h}_{ij} \nonumber\\
&&- \frac{H}{2}(\dot{h}_{ij} - \delta_{ij}\dot{h}_{kk}) -\left( H^2 + \dot{H} \right)h_{ij} - H^2\delta_{ij}h_{kk} - H\delta_{ij}\partial_k h_{k0} \nonumber\\
&&- \frac{1}{2}(\partial_i\dot{h}_{j0} + \partial_j\dot{h}_{i0}) - a^2 \delta_{ij}\delta\Lambda \, . \label{gij_pert}
\end{eqnarray}
\end{subequations}

As we mentioned in Section~\ref{emug}, due to the presence of $J_{\nu}$ the (non) conservation of the energy--momentum tensor~\eqref{em_vio} at first order leads to
\begin{equation}\label{gen_emt_pert}
    \partial_{\mu}\delta T^{\mu}_{\ \ \nu} + \bar{\Gamma}^{\mu}_{\mu \alpha}\delta T^{\alpha}_{\ \ \nu} + \delta \Gamma^{\mu}_{\mu \alpha}\bar{T}^{\alpha}_{\ \ \nu} - \bar{\Gamma}^{\alpha}_{\mu \nu}\delta T^{\mu}_{\ \ \alpha} - \delta \Gamma^{\alpha}_{\mu \nu}\bar{T}^{\mu}_{\ \ \alpha} = \frac{\delta J_{\nu}}{\kappa^2}\, .
\end{equation}

We can simplify these expressions by decomposing the perturbations into scalars, divergenceless vectors, and divergenceless traceless symmetric tensors. The perturbation of the metric $h_{\mu \nu}$ can be written as
\begin{subequations}\label{hwog}
\begin{eqnarray}
h_{00} &=& -E\, , \\
h_{i0} &=& a\left( \partial_i F + G_i \right)\, , \\
h_{ij} &=& a^2( A\delta_{ij} + \partial_i\partial_j B + \partial_j C_i + \partial_i C_j + D_{ij})\, ,
\end{eqnarray}
\end{subequations}
where $A\, , B\, , E\, , F$ are scalar perturbations, $C_i$ and $G_i$ are vector perturbations, and $D_{ij}$ are tensor perturbations. Particularly, $C_i\, , G_i$ and $D_{ij}$ satisfy
\begin{equation}
\partial_i C_i = \partial_i G_i = 0\, , \quad \partial_i D_{ij} = 0\, , \quad D_{ii} = 0\, .
\label{metricpertcond}
\end{equation}

Analogously, the energy--momentum tensor can be decomposed in a similar way, this is, we can rewrite Eq.\eqref{tpert} as
\begin{subequations}\label{dtwog}
\begin{eqnarray}
\delta T_{00} &=& -\bar{\rho} h_{00} + \delta \rho\, , \label{t00pertdc} \\
\delta T_{i0} &=& \bar{p}h_{i0} - (\bar{\rho} + \bar{p})(\partial_i v + \delta v_i^V)\, , \label{t0ipertdc} \\
\delta T_{ij} &=& \bar{p}h_{ij} + a^2\left( \delta_{ij}\delta p + \partial_i\partial_j \pi^S + \partial_i \pi_j^V + \partial_j\pi_i^V + \pi_{ij}^T \right)\, , \label{tijpertdc}
\end{eqnarray}
\end{subequations}
where we have decomposed the spatial part of the four--velocity perturbation as $v_i \equiv \partial_i v + \delta v_i^V$, with $\partial_i v$ the gradient of a scalar velocity potential, and $\delta v_i^V$ a divergenceless vector. The terms $\pi^S\, , \pi^V\, ,$ and $\pi^T$ represent dissipative corrections to the perturbation of the inertia tensor $\delta T_{ij}$. This quantities satisfy similar conditions to that of Eq.~\eqref{metricpertcond}
\begin{equation}
\partial_i\pi_i^V = \partial_i\delta v_i^V = 0\, , \quad \partial_i\pi_{ij}^T = 0\, , \quad \pi_{ii}^T=0\, .
\label{emtpertcond}
\end{equation}

Besides, the mixed components of the energy--momentum tensor~\eqref{mix_ind_emt} are given by
\begin{subequations}
\begin{eqnarray}
\delta T^{0}_{\ \ 0} &=& -\delta \rho\, , \\
\delta T^{i}_{\ \ 0} &=& a^{-2}(\bar{\rho} + \bar{p})(a\partial_i F + aG_i - \partial_i v - \delta v_i^V)\, , \\
\delta T^{0}_{\ \ i} &=& (\bar{\rho} + \bar{p})(\partial_i v + \delta v_i^V)\, , \\
\delta T^{i}_{\ \ j} &=& \delta_{ij}\delta p + \partial_i\partial_j\pi^S + \partial_i\pi_j^V + \partial_j\pi_i^V + \pi_{ij}^T\, , \\
\delta T &=& 3\delta p - \delta \rho + \nabla^2\pi^S\, .
\end{eqnarray}
\end{subequations}

As is the case in GR, in the linear regime of small fluctuations it is possible to separate the perturbations into three classes: scalar modes, vector modes, and tensor modes, which at linear order are completely independent from each other. We will be focused in the scalar modes of perturbations, and then, Eq.~\eqref{g00pert} is given by
\begin{eqnarray}
\kappa^2(\delta \rho + 3\delta p + \nabla^2\pi^S) &=& \frac{\nabla^2E}{a^2} + 3H\dot{E} + \frac{2}{a}\nabla^2\dot{F} + \frac{2H}{a}\nabla^2F - 3\ddot{A} - 6H\dot{A} + 6(H^2+\dot{H})E \nonumber \\
&&- 2H\nabla^2\dot{B}  - \nabla^2\ddot{B}+ 2\delta \Lambda\, ,
\label{g00pertsimp}
\end{eqnarray}
while Eq.~\eqref{gi0pert} gives
\begin{equation}
-\kappa^2(\bar{\rho} + \bar{p})\partial_i v = H\partial_i E - \partial_i \dot{A}\, ,
\label{gi0pertsimp}
\end{equation}
which is exactly the same as that of GR. Eq.~\eqref{gij_pert} can be separated in two parts: that proportional to $\delta_{jk}$, and that proportional to $\partial_j\partial_k$, which gives
\begin{subequations}\label{gijpertsimp}
\begin{eqnarray}
\kappa^2(\delta \rho - \delta p - \nabla^2\pi^S) &=& -H\dot{E} - 2(3H^2 + \dot{H})E - \frac{\nabla^2A}{a^2} + \ddot{A} + 6H\dot{A} + H\nabla^2\dot{B} - 2\frac{H}{a}\nabla^2F -\delta\Lambda\, ,\nonumber\\
\\
0 &=& \partial_i\partial_j(2\kappa^2a^2\pi^S + E + A - a^2\ddot{B} - 3a\dot{a}\dot{B} + 2a\dot{F} + 4\dot{a}F)\, .
\end{eqnarray}
\end{subequations}

On the other hand, the (non) conservation of the energy--momentum tensor (see Section~\ref{emug}) given by Eq.~\eqref{gen_emt_pert} will be now written as
\begin{subequations}\label{dtpertsimp}
\begin{eqnarray}
-\frac{\delta J_{0}}{\kappa^2} &=& \delta \dot{\rho} + 3H(\delta \rho + \delta p) + \nabla^2\left[ \frac{(\bar{\rho} + \bar{p})}{a}\left( \frac{v}{a} - F \right) + H\pi^S \right] + \frac{(\bar{\rho} + \bar{p})}{2}(3\dot{A} + \nabla^2\dot{B})\, , \nonumber \\
\\
\frac{\partial_i\delta J^S}{\kappa^2} &=& \partial_i\left\lbrace \delta p + \nabla^2\pi^S + \partial_t\left[ (\bar{\rho} + \bar{p})v \right] + 3H(\bar{\rho} + \bar{p})v + \frac{(\bar{\rho} + \bar{p})}{2}E \right\rbrace \, ,\label{dtpertsimp_momentum}
\end{eqnarray}
\end{subequations}
where we have decomposed the energy--momentum current violation perturbation in the same way as the other perturbed quantities, i.e., $\delta J_{\mu} = (\delta J_0\, , \delta J_i)$, and $\delta J_i = \partial_i\delta J^S + \delta J_i^V$ with $\partial_i\delta J_i^V=0$. We consider only the scalar modes $\delta J_0$ and $\delta J^S$.

The unimodular constraint at linear regime of perturbations on the determinant of the metric given by Eq.~\eqref{vol_diff_pert} can be written as
\begin{equation}
    3A + \nabla^2B + E = 0\, ,
    \label{uni_cons_pert}
\end{equation}
which coincides with that reported by~\cite{Gao:2014nia,Basak:2015swx} in their respective notations.

Notice that the Ricci scalar~\eqref{ricci_sc} is then written as
\begin{eqnarray}
     R &=& 6\left[ \left( \frac{\dot{a}}{a} \right)^2 + \frac{\ddot{a}}{a} \right] - 6\left[ \left( \frac{\dot{a}}{a} \right)^2 + \frac{\ddot{a}}{a} \right]E - 3\frac{\dot{a}}{a}\dot{E} - \frac{\nabla^2E}{a^2} - \frac{6}{a}\frac{\dot{a}}{a}\nabla^2F - \frac{2}{a}\nabla^2\dot{F} - \frac{2}{a^2}\nabla^2A\nonumber \\
     && +4\frac{\dot{a}}{a}\left( 3\dot{A} + \nabla^2\dot{B} \right) + 3\ddot{A} + \nabla^2\ddot{B}\, ,
\end{eqnarray}
and the perturbed energy--momentum current violation $\delta J_{\mu} = (1/4)\partial_{\mu}(\delta R + \kappa^2\delta T)$ is given by
\begin{eqnarray}
     \delta J_{\mu} &=& \frac{1}{4}\partial_{\mu}\left\lbrace - 6\left[ \left( \frac{\dot{a}}{a} \right)^2 + \frac{\ddot{a}}{a} \right]E - 3\frac{\dot{a}}{a}\dot{E} - \frac{\nabla^2E}{a^2} - \frac{6}{a}\frac{\dot{a}}{a}\nabla^2F - \frac{2}{a}\nabla^2\dot{F} - \frac{2}{a^2}\nabla^2A \right. \nonumber \\
     &&\left. + 4\frac{\dot{a}}{a}\left( 3\dot{A} + \nabla^2\dot{B} \right) 
     + 3\ddot{A} + \nabla^2\ddot{B} + \kappa^2\left( 3\delta p - \delta\rho \right)\right\rbrace\, .
     \label{jmupert}
\end{eqnarray}

From the above expression we notice that, besides the fact that the perturbed energy--momentum current violation $\delta J_{\mu}$ is function of the scalar metric perturbations $A\, , B\, , E\, , F\, ,$ as well as of the perturbed matter quantities $\delta\rho$ and $\delta p$, it is obtained that
\begin{equation}
    \delta J_0 = \partial_0\delta J^S\, .
    \label{j0js}
\end{equation}

The set of equations \eqref{g00pertsimp}--\eqref{uni_cons_pert} constitute the relativistic linear perturbations equations to describe the evolution of small fluctuations of a perfect fluid in an expanding universe within the framework of UG.

\section{Fixing the gauge}\label{gauge_fixing_UG}

\qquad The theory of General Relativity is invariant under diffeomorphism, which means that the equations will remain the same under general coordinate transformations. On the other hand, we have set in the previous Section that the geometry will be described by the sum of two metric tensors: one describing the background spacetime $\bar{g}_{\mu \nu}$ and which we have fixed to be the FLRW \eqref{flrw}, and the other metric $h_{\mu \nu}$ representing the small perturbations of the spacetime. Then, since the theory is invariant under diffeomorphism, and the metric of the background $g_{\mu \nu}$ is fixed, the components of the metric tensor for the perturbations $h_{\mu \nu}$ are not unique. In other words, we can choose how to fix the perturbations of the metric.

Consider the following coordinate transformation
\begin{equation}
x^{\mu} \rightarrow x^{\prime \mu} = x^{\mu} + \epsilon^{\mu}(x)\, ,
\label{ct}
\end{equation}
with $\epsilon^{\mu}(x)$ a small quantity as the other perturbations $h_{\mu \nu}\, , \delta \rho\, ,$ etc., and primes $(\ ^{\prime}\ )$ labels change of coordinates. Whereas in GR the 4--vector $\epsilon^{\mu}=(\epsilon^0\, , \epsilon^i)$ is arbitrary, in the case of UG it satisfies
\begin{equation}
    \nabla_{\mu}\epsilon^{\mu} = 0\, ,
\end{equation}
which reflects the rigidity of the spacetime volume under the unimodular condition. Notice that $\epsilon^0=-\epsilon_0$ and $\epsilon^i = a^{-2}\epsilon_i$. Developing the above expression we have
\begin{equation}
    \nabla_{\mu}\epsilon^{\mu} = \partial_{\mu}\epsilon^{\mu} + \bar{\Gamma}^{\mu}_{\mu \nu}\epsilon^{\nu}= \dot{\epsilon}_0 - \frac{\partial_i\epsilon_i}{a^2} +  3H\epsilon_0 = 0\, .
    \label{coord_trans_ug}
\end{equation}

Additionally, with the coordinate transformation \eqref{ct} the metric will transform as
\begin{equation}
g^{\prime}_{\mu \nu}(x^{\prime}) = g_{\lambda \kappa}\frac{\partial x^{\lambda}}{\partial x^{\prime \mu}}\frac{\partial x^{\kappa}}{\partial x^{\prime \nu}}\, .
\label{mt}
\end{equation}

Since we are in a scenario in which only the perturbed metric will be affected by a coordinate transformation (the unperturbed metric is given by the FRW line element), we implement \textit{gauge transformations} and we will attribute the whole change in $g_{\mu \nu}$ to a change in $h_{\mu \nu}$. Therefore, any change of coordinates $\Delta h_{\mu \nu}(x)$ on the perturbation of the metric of the form $h_{\mu \nu}(x) \rightarrow h_{\mu \nu}(x) + \Delta h_{\mu \nu}(x)$ must leaves invariant the field equations\footnote{This is the gravitational analogue to the electromagnetic potentials $\varphi$ and $\vec{A}$, which under gauge transformations, both fields the electric $\vec{E}$ and magnetic $\vec{B}$ remain the same, leaving invariant the Maxwell equations.}.
The change on the perturbation is defined as follows
\begin{equation}
\Delta h_{\mu \nu}(x) \equiv g^{\prime}_{\mu \nu}(x) - g_{\mu \nu}(x)\, ,
\end{equation}
which written in terms of its components once inserted Eq.~\eqref{mt}, and after expanding up to first order in perturbations, it can be shown that it is obtained
\begin{subequations}\label{hg}
\begin{eqnarray}
\Delta h_{00} &=& -2\dot{\epsilon}_0\, , \label{h00g} \\
\Delta h_{i0} &=& -\dot{\epsilon}_i - \partial_i \epsilon_0 + 2H\epsilon_i\, , \label{hi0g} \\
\Delta h_{ij} &=& - \partial_i \epsilon_j -\partial_j \epsilon_i + 2a\dot{a}\delta_{ij}\epsilon_0\, . \label{hijg}
\end{eqnarray}
\end{subequations}

Analogous to $\Delta h_{\mu \nu}(x)$, the change on the perturbation of the energy--momentum tensor will be
\begin{subequations}\label{dtg}
\begin{eqnarray}
\Delta \delta T_{00} &=& 2\bar{\rho}  \dot{\epsilon}_0 + \dot{\bar{\rho}}\epsilon_0\, , \label{dt00g} \\
\Delta \delta T_{i0} &=& -\bar{p}\dot{\epsilon}_i + \bar{\rho} \partial_i \epsilon_0 + 2\bar{p}H\epsilon_i\, , \label{dt0ig} \\
\Delta \delta T_{ij} &=& -\bar{p}\left( \partial_i \epsilon_j + \partial_j \epsilon_i \right) + \frac{\partial}{\partial t}(a^2\bar{p})\delta_{ij}\epsilon_0\, . \label{dtijg}
\end{eqnarray}
\end{subequations}

Following the same procedure, but this time applied to the energy--momentum current violation, we have
\begin{equation}
    \Delta \delta J_{\mu}(x) = -\bar{J}_{\lambda}(x)\frac{\partial \epsilon^{\lambda}}{\partial x^{\mu}} - \frac{\partial \bar{J}_{\mu}}{\partial x^{\lambda}}\epsilon^{\lambda}\, ,
\end{equation}
whose components are given by
\begin{subequations}\label{gauge_emcv}
\begin{eqnarray}
    \Delta \delta J_0 &=& 2\bar{J}_0\dot{\epsilon}_0 + \frac{\bar{J}_i}{a^2}\left( 2H\epsilon_i - \dot{\epsilon}_i \right)\, .\\
    \Delta \delta J_i &=& \bar{J}_0\partial_i\epsilon_0 + \dot{\bar{J}}_i\epsilon_0 - a^{-2}\left( \bar{J}_j\partial_i\epsilon_j + \epsilon_j\partial_j \bar{J}_i  \right)\, .
\end{eqnarray}
\end{subequations}

Since we have chosen comoving observers for the background (see the four--velocity in Eq.~\eqref{rpu_co}), it can be shown that the energy--momentum current violation is given by
\begin{equation}
    \bar{J}_{\mu} = \frac{1}{4}\nabla_{\mu}\left( \bar{R} + \kappa^2\bar{T} \right) = \frac{1}{4}\nabla_{\mu}\left[ 4\bar{\Lambda}(t) \right]\quad \Rightarrow\quad \bar{J}_{\mu} = \left[ \dot{\bar{\Lambda}}(t)\, , 0\, , 0\, , 0 \right]\, ,
\end{equation}
and then, Eq.~\eqref{gauge_emcv} gets reduced in the following simpler form
\begin{equation}
    \Delta \delta J_0 = 2\bar{J}_0\dot{\epsilon}_0\, ,\quad \Delta \delta J_i = \bar{J}_0\partial_i\epsilon_0\, .
    \label{gauge_trans_J}
\end{equation}

To be able to classify these gauge transformation into scalar, vector and tensor components, let us decompose the spatial part of $\epsilon^{\mu}$ into the gradient of a scalar $\epsilon^S$ and a divergenceless vector $\epsilon_i^V$ as follows
\begin{equation}
\epsilon_i = \partial_i \epsilon^S + \epsilon_i^V\, , \quad \text{with}\quad \partial_i\epsilon_i^V = 0\, ,
\end{equation}
and then, from Eq.~\eqref{coord_trans_ug} we obtain
\begin{equation}\label{ug_coord_trans_sca_mod}
    \dot{\epsilon}_0 - \frac{\nabla^2\epsilon^S}{a^2} +  3H\epsilon_0 = 0\, .
\end{equation}

Therefore, Eq.~\eqref{hg} and \eqref{dtg} give the gauge transformations of the metric components \eqref{hwog} and energy--momentum tensor \eqref{dtwog} respectively, and the scalar modes of the coordinate transformation obey~\eqref{ug_coord_trans_sca_mod}. For the metric perturbation we have
\begin{eqnarray}\label{gtforh}
\Delta A = \frac{2\dot{a}}{a}\epsilon_0\, &,& \quad \Delta B = -\frac{2}{a^2}\epsilon^S\, , \quad \Delta C_i = -\frac{1}{a^2}\epsilon_i^V\, , \quad \Delta D_{ij} = 0\, , \quad \Delta E = 2\dot{\epsilon}_0\, , \nonumber \\
\\
\Delta F &=& \frac{1}{a}\left( -\epsilon_0 - \dot{\epsilon}^S + \frac{2\dot{a}}{a}\epsilon^S \right)\, , \quad \Delta G_i = \frac{1}{a}\left( -\dot{\epsilon}_i^V + \frac{2\dot{a}}{a}\epsilon_i^V \right)\, , \nonumber
\end{eqnarray}
while for the energy--momentum tensor, the gauge transformations are given by
\begin{equation}
\Delta \delta \rho = \dot{\bar{\rho}}\epsilon_0\, , \quad \Delta \delta p = \dot{\bar{p}}\epsilon_0\, , \quad \Delta v = -\epsilon_0\, ,
\label{gtforemt}
\end{equation}
and the other terms are gauge invariants, this is
\begin{equation}
\Delta \pi^S = \Delta \pi_i^V = \Delta_{ij}^T = \Delta \delta u_i^V = 0\, .
\end{equation}

With expressions \eqref{gtforh} and \eqref{gtforemt} we can \textit{fix the gauge}, this is, we can choose particular values of the components of $\epsilon^{\mu}(x)$ to close the system of equations unambiguously. As was said before, our interest lies in the scalar perturbations, in which case the most general line element is written as
\begin{equation}\label{lineelem}
    ds^2 = -(1+E)dt^2 + 2a\partial_iFdtdx^i + a^2\left[ (1+A)\delta_{ij} + \partial_i\partial_j B \right]dx^idx^j\, ,
\end{equation}
and there are several choices we can consider to fix them. We want to analyze two gauges that are broadly used in the literature for cosmological perturbations: \textit{Newtonian gauge} and \textit{Synchronous gauge}. For a detailed study of these gauges in GR see~\cite{Ma:1994dv}.

\subsection{``Newtonian'' gauge: $B^{\prime}=0$ and $F^{\prime}=0$}

\qquad The gravitational potentials $E\, , F\, , A\, , B$ in~\eqref{lineelem} are general non--null solutions of the perturbed cosmological equations~\eqref{g00pertsimp},\eqref{gi0pertsimp},\eqref{gijpertsimp}, and~\eqref{dtpertsimp}. In this gauge, also known as conformal/longitudinal gauge, we choose $\epsilon^S$ such that $B^{\prime}=0$ and then ask for $\epsilon_0$ such that $F^{\prime}=0$, where primed quantities label gravitational potentials in the new coordinates. In first place, let us show that it is possible to fix unambiguously this gauge, i.e., the scalar components of the vector $\epsilon^{\mu}(x)$ given by $\epsilon_0$ and $\epsilon^S$ are equal to zero after performing coordinate transformations once the conditions requested above are satisfied. From~\eqref{gtforh} we have
\begin{subequations}\label{fngt}
    \begin{eqnarray}
         \Delta B &=& B^{\prime}-B = -\frac{2}{a^2}\epsilon^S\, ,\label{DB}\\
         \Delta F &=& F^{\prime}-F = \frac{1}{a}\left( -\epsilon_0 - \dot{\epsilon}^S + \frac{2\dot{a}}{a}\epsilon^S \right)\, .\label{DF}
    \end{eqnarray}
\end{subequations}

Solving for $\epsilon_0$ and $\epsilon^S$ from~\eqref{fngt} we obtain that the conditions $B^{\prime}=0$ and $F^{\prime}=0$ are satisfied when
\begin{equation}
\epsilon^S(t,\vec{x}) = \frac{1}{2}a^2(t)B(t,\vec{x})\, , \quad \epsilon_0(t,\vec{x}) = a(t)F(t,\vec{x}) + \frac{1}{2}a^2(t)\dot{B}(t,\vec{x})\, .
\label{coord_fix_ng}
\end{equation}

Now, by performing a new coordinate transformation $\tilde{\epsilon}^{\mu}(x)$, and requiring to remain in the Newtonian gauge, this is, choosing $\tilde{\epsilon}^S$ such that $B^{\prime \prime}=0$ and $\tilde{\epsilon}_0$ such that $F^{\prime \prime}=0$, we have
\begin{subequations}
    \begin{eqnarray}
         \Delta B &=& B^{\prime \prime}-B^{\prime} = -\frac{2}{a^2}\tilde{\epsilon}^S\, ,\label{DB2}\\
         \Delta F &=& F^{\prime \prime}-F^{\prime} = \frac{1}{a}\left( -\tilde{\epsilon}_0 - \dot{\tilde{\epsilon}}^S + \frac{2\dot{a}}{a}\tilde{\epsilon}^S \right)\, ,\label{DF2}
    \end{eqnarray}
\end{subequations}
where, since we already have that $B^{\prime}=F^{\prime}=0$, there is no other possible choice of coordinate transformation than $\tilde{\epsilon}_0 = 0$ and $\tilde{\epsilon}^S = 0$ and then, the remaining variables are totally determined. Therefore, from~\eqref{gtforh} the scalar gravitational potentials satisfy:
\begin{subequations}
    \begin{eqnarray}\label{gtforh_new}
\Delta A = A^{\prime \prime}-A^{\prime} = \frac{2\dot{a}}{a}\tilde{\epsilon}_0 = 0 \quad &\Rightarrow& \quad A^{\prime \prime} = A^{\prime} \neq 0\, ,\\
\Delta B = B^{\prime \prime}-B^{\prime} =-\frac{2}{a^2}\tilde{\epsilon}^S = 0 \quad &\Rightarrow& \quad B^{\prime \prime} = B^{\prime} = 0\, ,\\
\Delta E = E^{\prime \prime}-E^{\prime} = 2\dot{\tilde{\epsilon}}_0 = 0 \quad &\Rightarrow& \quad E^{\prime \prime} = E^{\prime} \neq 0\, ,\\
\Delta F = F^{\prime \prime}-F^{\prime} = \frac{1}{a}\left( -\tilde{\epsilon}_0 - \dot{\tilde{\epsilon}}^S + \frac{2\dot{a}}{a}\tilde{\epsilon}^S \right)= 0 \quad &\Rightarrow& \quad F^{\prime \prime} = F^{\prime} = 0\, ,
\end{eqnarray}
\end{subequations}
and then, the only non--null gravitational potentials in this gauge are $A$ and $E$. Thus, this gauge is completely fixed, and there is no remaining freedom to make any additional transformation. The unimodular condition~\eqref{uni_cons_pert} was not necessary to be implemented in order to fix this gauge. However, it will be taken into account in the equations of motion.

\subsection{``Synchronous'' gauge: $E^{\prime}=0$ and $F^{\prime}=0$}\label{syncgauge}

\qquad The choice for this gauge consists in fixing $\epsilon_0$ such that $E^{\prime}=0$, and then choose $\epsilon^S$ such that $F^{\prime}=0$. Using~\eqref{gtforh} we have
\begin{subequations}\label{fsgt}
    \begin{eqnarray}
         \Delta E &=& E^{\prime} - E = 2\dot{\epsilon}_0\, ,\label{DE}\\
         \Delta F &=& F^{\prime}-F = \frac{1}{a}\left( -\epsilon_0 - \dot{\epsilon}^S + \frac{2\dot{a}}{a}\epsilon^S \right)\, ,\label{DFs}
    \end{eqnarray}
\end{subequations}
from where it is obtained
\begin{equation}
\epsilon_0(t,\vec{x}) = f_1(\vec{x}) - \frac{1}{2}\int E(t,\vec{x})dt\, , \quad \epsilon^S(t,\vec{x}) = a^2(t)\left[f_2(\vec{x}) - \int \frac{a(t)F(t,\vec{x})+\epsilon_0(t,\vec{x})}{a^2(t)}\right]\, .
\label{coord_fix_sg}
\end{equation}

Now, considering a new coordinate transformation $\tilde{\epsilon}^{\mu}(x)$, and again requiring to remain in the synchronous gauge choosing $\tilde{\epsilon}_0$ such that $E^{\prime \prime}=0$ and $\tilde{\epsilon}^S$ such that $F^{\prime \prime}=0$, we have
\begin{subequations}
    \begin{eqnarray}
         \Delta E &=& E^{\prime \prime}-E^{\prime} = 2\dot{\tilde{\epsilon}}_0\, ,\label{DE2}\\
         \Delta F &=& F^{\prime \prime}-F^{\prime} = \frac{1}{a}\left( -\tilde{\epsilon}_0 - \dot{\tilde{\epsilon}}^S + \frac{2\dot{a}}{a}\tilde{\epsilon}^S \right)\, .\label{DF2s}
    \end{eqnarray}
\end{subequations}

This time, it it found that
\begin{equation}
\tilde{\epsilon}_0(\vec{x}) = f_3(\vec{x})\, , \quad \tilde{\epsilon}^S(t,\vec{x}) = a^2(t)\left[f_4(\vec{x}) - \tilde{\epsilon}_0(\vec{x})\int \frac{dt}{a^2(t)}\right]\, .
\label{coord_fix_sg_2}
\end{equation}

In order to completely fix the synchronous gauge, we have to determine in some way the arbitrary scalar functions $f_3$ and $f_4$. We can perform a new coordinate transformation, but it can be proved that successive gauge transformations lead to the same mathematical structure of $(\tilde{\epsilon}_0, \tilde{\epsilon}^S)$, with a new couple of spatial functions. For instance, it can be shown that for a third gauge transformation $(\tilde{\tilde{\epsilon}}_0,\tilde{\tilde{\epsilon}}^S)$ it is possible to obtain
\begin{equation}
\tilde{\tilde{\epsilon}}_0(\vec{x}) = f_5(\vec{x})\, , \quad \tilde{\tilde{\epsilon}}^S(t,\vec{x}) = a^2(t)\left[f_6(\vec{x}) - \tilde{\tilde{\epsilon}}_0(\vec{x})\int \frac{dt}{a^2(t)}\right]\, .
\label{coord_fix_sg_3}
\end{equation}

Therefore, we are always left with two arbitrary spatial functions, and the synchronous gauge remains ambiguous. Notwithstanding, all the spatial functions in $\tilde{\epsilon}^S, \tilde{\tilde{\epsilon}}^S$ and so on, affects only the initial coordinate labelling, and it is only the spatial function in the time components $\tilde{\epsilon}_0, \tilde{\tilde{\epsilon}}_0$, etc, which remains as a spurious degree of freedom, and it will have repercussions on physical quantities if it is not properly determined~\cite{Malik:2008im}. We can safely keep then the coordinate transformations~\eqref{coord_fix_sg_2} and~\eqref{coord_fix_sg_3} as
\begin{equation}
\tilde{\epsilon}_0(\vec{x}) = f_3(\vec{x})\, , \quad \tilde{\epsilon}^S(t,\vec{x}) = -a^2(t) \tilde{\epsilon}_0(\vec{x})\int \frac{dt}{a^2(t)}\, ,
\label{coord_fix_sg_2_red}
\end{equation}
\begin{equation}
\tilde{\tilde{\epsilon}}_0(\vec{x}) = f_5(\vec{x})\, , \quad \tilde{\tilde{\epsilon}}^S(t,\vec{x}) = -a^2(t) \tilde{\tilde{\epsilon}}_0(\vec{x})\int \frac{dt}{a^2(t)}\, ,
\label{coord_fix_sg_3_red}
\end{equation}
respectively, and so on for successive coordinate transformations. Then, we have to deal with only one arbitrary function. Notice that if $f_5=0$, then both $\tilde{\tilde{\epsilon}}_0=0$ and $\tilde{\tilde{\epsilon}}^S=0$, and the gauge is completely fixed. The standard approach in GR to handle this situation of ambiguity in the coordinates, is to move to the CDM frame of reference, i.e., by choosing a coordinate transformation comoving with the CDM fluid. From Eq.~\eqref{dtpertsimp_momentum}, and for the GR case where $\delta J^S=0$, we have
\begin{equation}
    \delta p^{\prime} + \nabla^2\pi^{S\ \prime} + \partial_t\left[ (\bar{\rho} + \bar{p})v^{\prime} \right] + 3H(\bar{\rho} + \bar{p})v^{\prime} + \frac{(\bar{\rho} + \bar{p})}{2}E^{\prime} = 0\, ,
\end{equation}
which for the synchronous gauge we already show that it is asked for new coordinates where $E^{\prime}=0$, in whose case we have
\begin{equation}
    \delta p^{\prime} + \nabla^2\pi^{S\ \prime} + \partial_t\left[ (\bar{\rho} + \bar{p})v^{\prime} \right] + 3H(\bar{\rho} + \bar{p})v^{\prime} = 0\, ,
\end{equation}
and when considering the CDM fluid, there is no pressure nor anisotropic stress. Then, the previous equation applied to CDM is
\begin{equation}
    \dot{\bar{\rho}}_{CDM}v^{\prime}_{CDM} + \bar{\rho}_{CDM}\dot{v^{\prime}}_{CDM} +3H\bar{\rho}_{CDM}v^{\prime}_{CDM} = 0\, .
    \label{em_cons_cdm_gr}
\end{equation}

In the case of GR, the energy--momentum conservation at background level for CDM is (see Eq.~\eqref{em_cons_bg} with $\bar{J}_0 = 0$),
\begin{equation}
    \dot{\bar{\rho}}_{CDM} = -3H\bar{\rho}_{CDM}\, ,
\end{equation}
where Eq.~\eqref{em_cons_cdm_gr} gets reduced to
\begin{equation}
     \bar{\rho}_{CDM}\dot{v^{\prime}}_{CDM} = 0\quad \Rightarrow \quad v^{\prime}_{CDM} = f(\vec{x})\, ,
    \label{v_cte_cdm_gr}
\end{equation}
and then, in GR the CDM peculiar velocity is a function of spatial coordinates only. This is a crucial result in order to completely fix the synchronous gauge in GR, because now we can consider the coordinate transformation for $v_{CDM}$ as follows: from~\eqref{gtforemt} we have
\begin{equation}
    \Delta v_{CDM} = v_{CDM}^{\prime} - v_{CDM} = -\epsilon_0\quad \Rightarrow\quad v_{CDM}^{\prime} = v_{CDM}  -\epsilon_0\, ,
\end{equation}
where $\epsilon_0$ is given by~\eqref{coord_fix_sg}. After a new change of coordinates, we have
\begin{equation}
    \Delta v_{CDM} = v_{CDM}^{\prime \prime} - v_{CDM}^{\prime} = -\tilde{\epsilon}_0\quad \Rightarrow\quad v_{CDM}^{\prime \prime} = v_{CDM}^{\prime}  -\tilde{\epsilon}_0\, ,
\end{equation}
but as we already see from Eq.~\eqref{coord_fix_sg_2} and~\eqref{v_cte_cdm_gr}, both functions $v_{CDM}^{\prime}$ and $\tilde{\epsilon}_0$ are functions only of spatial coordinates. Thus, we can choose $\tilde{\epsilon}_0 = v_{CDM}^{\prime}$ in order to have $v_{CDM}^{\prime \prime} = 0$, and then we will be in a reference frame comoving with the CDM fluid. Now, under a new change of coordinates and asking for remaining in the CDM fluid reference frame, which is given by the condition $v_{CDM}^{\prime \prime \prime}=0$, we obtain
\begin{equation}
    \Delta v_{CDM} = v_{CDM}^{\prime \prime \prime} - v_{CDM}^{\prime \prime} = -\tilde{\tilde{\epsilon}}_0\quad \Rightarrow\quad v_{CDM}^{\prime \prime \prime} = v_{CDM}^{\prime \prime}  -\tilde{\tilde{\epsilon}}_0\, ,
\end{equation}
but since $v_{CDM}^{\prime \prime \prime} = v_{CDM}^{\prime \prime} = 0$, there is no other possible choice for $\tilde{\tilde{\epsilon}}_0$ that $\tilde{\tilde{\epsilon}}_0=0$. This completely fix the synchronous gauge in GR, as can be seen from Eq.~\eqref{coord_fix_sg_3_red}.

Whereas in GR this gauge is fixed as we have shown above, in UG the scenario is completely different. Basically, the issue we have to deal with is based on the fact that it is not possible to move to the CDM reference frame due to the energy--momentum current violation $J_{\mu}$. From Eq.~\eqref{dtpertsimp_momentum}, we have
\begin{equation}
    \delta p^{\prime} + \nabla^2\pi^{S\ \prime} + \partial_t\left[ (\bar{\rho} + \bar{p})v^{\prime} \right] + 3H(\bar{\rho} + \bar{p})v^{\prime} + \frac{(\bar{\rho} + \bar{p})}{2}E^{\prime} = \frac{\delta J^{S\ \prime}}{\kappa^2}\, .
\end{equation}

In the synchronous gauge, and for the CDM fluid, the previous equation is written as
\begin{equation}
    \dot{\bar{\rho}}_{CDM}v^{\prime}_{CDM} + \bar{\rho}_{CDM}\dot{v^{\prime}}_{CDM} +3H\bar{\rho}_{CDM}v^{\prime}_{CDM} = \frac{\delta J^{S\ \prime}}{\kappa^2}\, ,
    \label{em_cons_cdm_gr}
\end{equation}
but now, in UG the energy--momentum conservation at background level for CDM is (see Eq.~\eqref{em_cons_bg}),
\begin{equation}
    \dot{\bar{\rho}}_{CDM} = -3H\bar{\rho}_{CDM} -\frac{\bar{J}_0(t)}{\kappa^2}\, ,
\end{equation}
which leads to
\begin{equation}
     \dot{v^{\prime}}_{CDM} = \frac{\bar{J}_0(t)v^{\prime}_{CDM} + \delta J^{S\ \prime}}{\kappa^2\bar{\rho}_{CDM}}\, ,
    \label{v_cte_cdm_ug}
\end{equation}
and then, in UG it is no longer true in general that $v^{\prime}_{CDM}$ is a function of spatial coordinates only. In fact, by solving Eq.~\eqref{v_cte_cdm_ug} we obtain
\begin{equation}
    v^{\prime}_{CDM} = g(\vec{x})e^{\frac{1}{\kappa^2}\int\frac{\bar{J}_0(t)}{\bar{\rho}_{CDM}(t)}dt}\left[1 + \frac{1}{\kappa^2g(\vec{x})}\int \frac{e^{-\frac{1}{\kappa^2}\int\frac{\bar{J}_0(t)}{\bar{\rho}_{CDM}(t)}dt}\delta J^{S\ \prime}}{\bar{\rho}_{CDM}(t)}dt \right]\, .
\end{equation}

Besides, notice that it is not possible to choose a coordinate system where there is not perturbations of the energy--momentum current violation, as we can see in Eq.~\eqref{gauge_trans_J} from the fact that we can not fix neither of $\epsilon_0, \tilde{\epsilon}_0, \tilde{\tilde{\epsilon}}_0$, etc, equal to zero. Moreover, we are left with an arbitrary function $g(\vec{x})$ that must be determined. Therefore, within the framework of UG, the cosmological perturbations in the synchronous gauge do not allow to choose the CDM comoving frame due to the presence of the energy--momentum current violation $\bar{J}_0$ and its scalar perturbation $\delta{J}^S$. Thus, this gauge can not be fixed in this way as in GR.

We can think in using the unimodular constraint for perturbations~\eqref{uni_cons_pert}: for the synchronous gauge it reads
\begin{equation}
    3A^{\prime \prime \prime} = -\nabla^2B^{\prime \prime \prime}\, .
\end{equation}

For the gauge transformations~\eqref{gtforh}, it is written as
\begin{equation}
\nabla^2\tilde{\tilde{\epsilon}}^S(t,\vec{x}) = 3a(t)\dot{a}(t)\tilde{\tilde{\epsilon}}_0(\vec{x})\, ,
\end{equation}
which from Eq.~\eqref{coord_fix_sg_3_red} we obtain,
\begin{equation}
    \left[\int\frac{dt}{a^2(t) }\right]\nabla^2f_5(\vec{x}) = -\left[\frac{3\dot{a}(t)}{a(t)}\right]f_5(\vec{x})\, .
    \label{gt_uc}
\end{equation}

The differential equation from above can be solved by the method of variable separation, and we have
\begin{equation}
    \frac{\nabla^2f_5(\vec{x})}{f_5(\vec{x})} = \mathcal{C}\, ,\quad -\left[\frac{3\dot{a}(t)}{a(t)\int dt/a^2(t)}\right]=\mathcal{C}\, ,
    \label{diff_eq_sep}
\end{equation}
with $\mathcal{C}$ a constant. The second expression is an integro--differential equation for the scale factor $a(t)$, which can be written as
\begin{equation}
     \frac{da(t)}{dt}=-\frac{\mathcal{C}a(t)}{3}\int\frac{ dt}{a^2(t)}\quad \Rightarrow\quad \dot{H} = -\frac{\mathcal{C}}{3a^2(t)} \, ,
     \label{int_diff_a}
\end{equation}
which demands a very particular solution for the background equations~\eqref{bg_eqs} that does not need to be a physical solution for the expansion of the universe filled with a particular matter content. Moreover, since the scale factor is solution of the background dynamics, it will not necessarily satisfy~\eqref{int_diff_a}. In fact, such equation is way to restrictive to govern the expansion of the universe. For instance, we can consider a general power law solution $a(t) = (t/t_0)^p$, where $t_0$ is the present day. It can be shown that this solution is valid only when $p=2$ and $\mathcal{C}=18/t_0^2$. However, this constitutes a very particular background evolution for the dynamics of any matter component, as we mentioned above. Doing the same for a late time solution within the UG framework (see Appendix B in~\cite{LinaresCedeno:2020uxx}), where the scale factor is given by $a(t) = \left[A\sinh^2(Bt)\right]^C$ with $A,B,C$ constants, we have that it is not possible to satisfy~\eqref{int_diff_a}.

On the other hand, the first expression in Eq.~\eqref{diff_eq_sep}, is the Poisson equation with source term given by the function itself,
\begin{equation}
    \nabla^2f_5(\vec{x}) = \mathcal{C}f_5(\vec{x})\, .
    \label{poisson}
\end{equation}

Thus, the only way to have a scale factor driven by the background dynamics, and simultaneously the unimodular condition being satisfied at the level of perturbations in this gauge~\eqref{gt_uc}, is through the trivial solution $f_5(\vec{x}) = 0$. But this is precisely what we need to fix the synchronous gauge, as can be seen from Eq.~\eqref{coord_fix_sg_3_red}. Therefore, when considering non--gravitational interactions it is not possible to consistently fix the synchronous gauge in UG by choosing the comoving frame of CDM. Instead, the unimodular condition appears to be useful to completely fix this gauge. This has serious repercussions on the possibility of implementing cosmological models based on UG in Boltzmann solvers such as \texttt{CAMB}~\cite{Lewis:1999bs} and \texttt{CLASS}~\cite{Lesgourgues:2011re}, as we will discuss later.

\subsection{Alternative gauge choice: $B^{\prime}=0$ and unimodular constraint}\label{alternative_gauge}

\qquad This is the approach implemented by~\cite{Gao:2014nia} with the aim of keeping two geometric degrees of freedom, just as in GR perturbations. Moreover, in such work choose $B^{\prime}=0$ in order to compare with the Newtonian gauge in GR. Then, the line element under this choice can be written as
\begin{equation}\label{lineelem_B}
    ds^2 = -(1-3A^{\prime})dt^2 + 2a\partial_iF^{\prime}dtdx^i + a^2(1+A^{\prime})\delta_{ij}dx^idx^j\, ,
\end{equation}
where the only degrees of freedom are $A^{\prime}$ and $F^{\prime}$. However, we will show that this choice does not determined unambiguously these gravitational potentials, and spurious effects due to this not properly fixed gauge will affect physical quantities such as energy density and pressure perturbations.

Similar to previous procedures, and as in the Newtonian gauge, we ask for $\epsilon^S=0$ such that $B^{\prime}=0$ (see Eq.~\eqref{DB} and~\eqref{coord_fix_ng}), this is
\begin{equation}
    \Delta B = B^{\prime}-B = -\frac{2}{a^2}\epsilon^S\quad \Rightarrow\quad \epsilon^S(t,\vec{x}) = \frac{a^2}{2}B(t,\vec{x})\, .
\end{equation}

Now, instead of asking for $\epsilon_0$ such that $F^{\prime}=0$, we follow the approach of~\cite{Gao:2014nia} by imposing the unimodular condition~\eqref{uni_cons_pert}, which in terms of the scalar components of the coordinate transformation $\epsilon^{\mu}$ is written as
\begin{equation}
    \dot{\epsilon_0} + 3\frac{\dot{a}}{a}\epsilon_0 = 0\quad\Rightarrow \quad \epsilon_0(t,\vec{x}) = \frac{f_1(\vec{x})}{a^3(t)}\, .
\end{equation}

It can be seen that we have left with an arbitrary spatial function $f_1(\vec{x})$. A new coordinate transformation $\tilde{\epsilon}^{\mu}$ where $B^{\prime \prime}=0$ leads to
\begin{equation}
    \Delta B = B^{\prime \prime}-B^{\prime} = -\frac{2}{a^2}\tilde{\epsilon}^S\quad \Rightarrow\quad \tilde{\epsilon}^S(t,\vec{x}) = 0\, ,
\end{equation}
and the remaining scalar component $\tilde{\epsilon}_0$ must be zero to completely fix the gauge. However, once the unimodular condition is imposed one more time, we have
\begin{equation}
    \dot{\tilde{\epsilon}}_0 + 3\frac{\dot{a}}{a}\tilde{\epsilon}_0 = 0\quad\Rightarrow \quad \tilde{\epsilon}_0(t,\vec{x}) = \frac{f_2(\vec{x})}{a^3(t)}\, ,
\end{equation}
and then, we still have an arbitrary spatial function $f_2(\vec{x})$. It can be shown that successive coordinates transformations lead to the same results, i.e., $\tilde{\tilde{\epsilon}}^S = 0$, but $\tilde{\tilde{\epsilon}}_0$ is always in terms of an arbitrary spatial function. This gauge freedom will affect not only the remaining gravitational potentials, which after such transformations are
\begin{subequations}\label{f_and_a}
    \begin{eqnarray}
        F^{\prime \prime}(t,\vec{x}) &=& F(t,\vec{x}) - \frac{a}{2}\dot{B}(t,\vec{x}) - \frac{f_3(\vec{x})}{a^4}\, , \\
        A^{\prime \prime}(t,\vec{x}) &=& A(t,\vec{x}) + 2\frac{\dot{a}}{a^4}f_3(\vec{x})\, ,
    \end{eqnarray}
\end{subequations}
with $f_3 = f_1+f_2$, but also to the energy density perturbation, which from Eq.~\eqref{gtforemt} is
\begin{equation}
    \delta \rho^{\prime \prime} = \delta \rho + \frac{\dot{\bar{\rho}}}{a^3(t)}f_3(\vec{x})\, ,
\end{equation}
and similarly for both, the pressure perturbation $\Delta \delta p$ and peculiar velocity $\Delta v$ transformations. Moreover, the Sachs--Wolfe effect~\cite{Sachs:1967er} has been derived in UG in~\cite{Gao:2014nia} in order to spot differences between GR and UG through possible signatures in the anisotropies of the CMB. In our notation, they obtain that
\begin{equation}
    \left( -\frac{3A^{\prime \prime}}{2} + \frac{\delta T}{\bar{T}} + a\dot{F}^{\prime \prime} \right) = {\rm{ctte}}\, ,
    \label{sw}
\end{equation}
but as it is shown in Eq.~\eqref{f_and_a}, both gravitational potentials are not completely determined due to the arbitrary spatial function $f_3(\vec{x})$. Therefore, even when the differences obtained in~\cite{Gao:2014nia} between GR and UG are negligible under the assumptions they considered, it is important to study physical observables such as the CMB radiation, with a proper gauge choice without spurious degrees of freedom. This will be discussed in detail in the next Section.

\section{Physical implications of cosmological perturbations in UG}\label{phys_imp}

\qquad Now that we have fixed both, the Newtonian and Synchronous gauge in UG, we are able to write down the dynamical equations for the linear perturbations in each of these gauges. As we will see below, the unimodular condition~\eqref{uni_cons_pert} will reduce the degrees of freedom from two to only one gravitational potential. It is possible to find solutions for the density contrasts in each of these gauges in terms of the corresponding metric scalar perturbation. Besides, we obtain proper derivation of the Sachs--Wolfe effect within the UG framework for the Newtonian gauge, and we show that there is only a modification in the coefficient of the gravitational potential derivative, but the physical result is exactly the same as that obtained in GR.

\subsection{Linear perturbations: Newtonian gauge}\label{lp_ng_sec}

\qquad The line element~\eqref{lineelem} is written as\footnote{In order to keep the notation simple, notice that we drop out the primes $(^{\prime})$ since we already know that it is possible to find a consistent coordinates transformation where $E$ and $A$ are the only physical degrees of freedom for the gravitational perturbations, prior to impose the unimodular constraint.},
\begin{equation}\label{lineelem_new}
    ds^2 = -(1+E)dt^2 + a^2 (1+A)\delta_{ij} + dx^idx^j\, ,
\end{equation}
but we will use the standard notation for $E$ and $A$ in this gauge, which is given by $E\equiv 2\Phi$, and $ A\equiv -2\Psi$, 
and then, the perturbed line element~\eqref{lineelem_new} takes the form
\begin{equation}\label{new_ds2}
    ds^2 = -(1+2\Phi)dt^2 + a^2 (1-2\Psi)\delta_{ij}dx^idx^j\, .
\end{equation}

The evolution of perturbations given by Eq.~\eqref{g00pertsimp}--\eqref{gijpertsimp} are then written as
\begin{subequations}\label{efeng}
\begin{eqnarray}
\frac{\kappa^2}{2}(\delta \rho + 3\delta p + \nabla^2\pi^S) &=& \frac{\nabla^2\Phi}{a^2} + 3H\dot{\Phi} + 3\ddot{\Psi} + 6H\dot{\Psi} + 6(H^2+\dot{H})\Phi + \delta\Lambda\, ,\\
-\frac{\kappa^2}{2}(\bar{\rho} + \bar{p})\partial_i v &=& H\partial_i\Phi + \partial_i\dot{\Psi}\, , \\
-\frac{\kappa^2}{2}(\delta \rho - \delta p - \nabla^2\pi^S) &=& H\dot{\Phi} + 2(3H^2 + \dot{H})\Phi - \frac{\nabla^2\Psi}{a^2} + \ddot{\Psi} + 6H\dot{\Psi} + \frac{\delta \Lambda}{2}\, , \label{poisGRpert}\\
\kappa^2a^2\partial_i\partial_j \pi^S &=& \partial_i\partial_j(\Psi - \Phi)\, , \label{phipsi}
\end{eqnarray}
\end{subequations}
and Eq.~\eqref{dtpertsimp} for the energy--momentum tensor becomes
\begin{subequations}\label{emtng}
\begin{eqnarray}
\dot{\delta \rho} + 3H(\delta \rho + \delta p) + \nabla^2\left[ \frac{(\bar{\rho} + \bar{p})}{a^2}v + H\pi^S \right] - 3(\bar{\rho} + \bar{p})\dot{\Psi} &=& -\frac{\delta J_0}{\kappa^2}\, , \label{ecng}\\
\delta p + \nabla^2\pi^S + \partial_t\left[ (\bar{\rho} + \bar{p})v \right] + 3H(\bar{\rho} + \bar{p})v + (\bar{\rho} + \bar{p})\Phi &=& \frac{\delta J^S}{\kappa^2}\, . \label{mcng}
\end{eqnarray}
\end{subequations}

We have 6 equations \eqref{efeng}--\eqref{emtng} and 6 variables to determine: $\delta \rho\, , \delta p\, , \pi^S\, \delta u\, , \Phi\, ,$ and $\Psi$. In particular, variables $\Phi$ and $\Psi$ differ by the scalar anisotropic term $\pi^S$ as can be seen from Eq.~\eqref{phipsi}. In the particular case of a perfect fluid without dissipative corrections, $\pi^S=0$, and we obtain that $\Phi = \Psi$. However, in UG we have that the choice of coordinates such that $B=F=0$ leads to the following unimodular constraint for perturbations~\eqref{uni_cons_pert}
\begin{equation}
    3\Psi = \Phi\, ,
    \label{uc_ng}
\end{equation}
and the previous equations acquire the form
\begin{subequations}
\begin{eqnarray}
\frac{\kappa^2}{6}(\delta \rho + 3\delta p + \nabla^2\pi^S) &=& \frac{\nabla^2\Psi}{a^2} + 4H\dot{\Psi} + \ddot{\Psi} + 6(H^2+\dot{H})\Psi + \frac{\delta\Lambda}{3}\, ,\\
-\frac{\kappa^2}{2}(\rho + p)\partial_i v &=& 3H\partial_i\Psi + \partial_i\dot{\Psi}\, , \\
-\frac{\kappa^2}{2}(\delta \rho - \delta p - \nabla^2\pi^S) &=& 6(3H^2 + \dot{H})\Psi - \frac{\nabla^2\Psi}{a^2} + \ddot{\Psi} + 9H\dot{\Psi} + \frac{\delta \Lambda}{2}\, , \label{pois_ug_pert}\\
-\frac{\kappa^2}{2}a^2\partial_i\partial_j \pi^S &=& \partial_i\partial_j \Psi\, , \label{phipsi_ug}\\
-\frac{\delta J_0}{\kappa^2} &=& \dot{\delta \rho} + 3H(\delta \rho + \delta p) + \nabla^2\left[ \frac{(\bar{\rho} + \bar{p})}{a^2}v + H\pi^S \right] - 3(\bar{\rho} + \bar{p})\dot{\Psi} \, , \label{ecng}\\
\frac{\delta J^S}{\kappa^2} &=& \delta p + \nabla^2\pi^S + \partial_t\left[ (\bar{\rho} + \bar{p})v \right] + 3H(\bar{\rho} + \bar{p})v + 3(\bar{\rho} + \bar{p})\Psi\, , \label{mcng}
\end{eqnarray}
\end{subequations}
and the perturbation for the energy--momentum current violation in this gauge is
\begin{eqnarray}
     \delta J_{\mu} &=& \frac{1}{4}\partial_{\mu}\left\lbrace - 36\left[ \left( \frac{\dot{a}}{a} \right)^2 + \frac{\ddot{a}}{a} \right]\Psi - 44\frac{\dot{a}}{a}\dot{\Psi} - \frac{2\nabla^2\Psi}{a^2} - 6\ddot{\Psi}  + \kappa^2\left( 3\delta p - \delta\rho \right)\right\rbrace\, .
     \label{jmupert}
\end{eqnarray}

Notice that we need the presence of the scalar anisotropic stress $\pi^S$ in order to have non--null gravitational potential $\Psi$, as can be seen from Eq.~\eqref{phipsi_ug}. In this sense, we can see that strictly the Newtonian gauge is not recovered in UG. This was already reported in previous literature, and recently by~\cite{Fabris:2021atr}. However, different from the approach of the mentioned work, we will keep the anisotropic stress term in order to find solutions for physical quantities, such as the CDM density contrast $\delta_{CDM}\equiv \delta \rho_{CDM}/\bar{\rho}_{CDM}$. We have then that $\bar{p}_{CDM} = \delta p_{CDM} = 0$, but different from the standard model in $\Lambda CDM$, the dark matter particle will have $\pi^S_{CDM}\neq 0$. Combining the previous equations, it can be shown that in a matter--dominated era the CDM density contrast in UG for the Newtonian gauge $\delta_{CDM(new)}^{UG}$ is given by
\begin{equation}
     \delta_{cdm(new)}^{{\rm{UG}}} = \left(\frac{4}{3}\right)\left(-\frac{2}{H}\dot{\Psi}\right) - 2\left[ 1 + \left(\frac{1}{6}\right)\left(\frac{k^2}{3a^2H^2}\right) \right]\left(3\right)\Psi\, ,
     \label{delta_cdm_ug_new}
\end{equation}
and it can be seen that it differs only by numerical factors from the GR result (see Eq.(12) in~\cite{Dent:2008ek}),
\begin{equation}
    \delta_{cdm}^{{\rm{GR}}} = -\frac{2}{H}\dot{\Psi} - 2\left(1 + \frac{k^2}{3a^2H^2} \right)\Psi\, ,
    \label{delta_cdm_gr_new}
\end{equation}
where we have used $\nabla \rightarrow -k^2$ for solutions in Fourier space. Thus, once the gravitational potential $\Psi$ is known, it is possible to follow the cosmological evolution of CDM fluctuations. Moreover, if a particular model for the non--gravitational interaction is considered at background level, such information will be in the Hubble parameter $H$ through the Friedmann equation~\eqref{friedmann}, and constraints could be put on cosmological models of UG by studying LSS of the universe through the \textit{Matter Power Spectrum} (MPS). Also notice that, even when neglecting the energy--momentum current violation, and GR is recovered at background level, the unimodular constraint at the level of linear perturbations changes the evolution of CDM fluctuations, as can be seen when comparing the coefficients of Eqs.~\eqref{delta_cdm_ug_new} and~\eqref{delta_cdm_gr_new}.

\subsection{Linear perturbations: Synchronous gauge}\label{lp_sg_sec}

In this case the perturbed line element is written as
\begin{equation}
    ds^2 = -dt^2 + a^2\left[ (1+A)\delta_{ij} + \partial_i\partial_j B \right]dx^idx^j\, ,
\end{equation}

The field equations~\eqref{g00pertsimp}--\eqref{gijpertsimp} in this gauge are given by
\begin{subequations}\label{efesg}
\begin{eqnarray}
\kappa^2(\delta \rho + 3\delta p + \nabla^2\pi^S) &=& -3\ddot{A} - 6H\dot{A} - \nabla^2\ddot{B} - 2H\nabla^2\dot{B} + 2\delta\Lambda\, ,\label{efeijsg}\\
\kappa^2(\bar{\rho} + \bar{p})v &=& \dot{A}\, ,\\
\kappa^2(\delta \rho - \delta p - \nabla^2\pi^S) &=& -\frac{\nabla^2A}{a^2} + \ddot{A} + 6H\dot{A} + H\nabla^2\dot{B} - \delta\Lambda\, , \\
2\kappa^2a^2\pi^S &=& -A + a^2\ddot{B} + 3a\dot{a}\dot{B}\, ,
\end{eqnarray}
\end{subequations}
and Eq.~\eqref{dtpertsimp} for the energy--momentum tensor now takes the form
\begin{subequations}\label{emtsg}
\begin{eqnarray}
\dot{\delta \rho} + 3H(\delta \rho + \delta p) + \nabla^2\left[ \frac{(\bar{\rho} + \bar{p})}{a^2}v + H\pi^S \right] + \frac{(\bar{\rho} + \bar{p})}{2}(3\dot{A} + \nabla^2\dot{B}) &=& -\frac{\delta J_0}{\kappa^2}\, .\label{ecsg}\\
\delta p + \nabla^2\pi^S + \partial_t\left[ (\bar{\rho} + \bar{p})v \right] + 3H(\bar{\rho} + \bar{p})v &=& \frac{\delta J^S}{\kappa^2}\, , \label{emtsgmom}
\end{eqnarray}
\end{subequations}
and the energy--momentum current violation perturbation~\eqref{jmupert} is given by
\begin{equation}
    \delta J_{\mu} = \frac{1}{4}\partial_{\mu}\left[ -\frac{2}{a^2}\nabla^2A + 4\frac{\dot{a}}{a}\left( 3\dot{A} + \nabla^2\dot{B} \right) + 3\ddot{A} + \nabla^2\ddot{B} + \kappa^2\left( 3\delta p - \delta\rho \right) \right]\, .
    \label{jmupert_syn}
\end{equation}

However, in UG we have that the choice of coordinates of Section~\ref{syncgauge} leads to the following unimodular constraint~\eqref{uni_cons_pert} for perturbations
\begin{equation}
    3A = -\nabla^2B\, ,
\end{equation}
and the previous equations are written as follows
\begin{subequations}
\begin{eqnarray}
     \kappa^2(\delta \rho + 3\delta p + \nabla^2\pi^S) &=&  2\delta\Lambda\, ,\label{efeijsg}\\
\kappa^2(\bar{\rho} + \bar{p})v &=& \dot{A}\, ,\\
\kappa^2(\delta \rho - \delta p - \nabla^2\pi^S) &=& -\frac{\nabla^2A}{a^2} + \ddot{A} + 3H\dot{A} - \delta\Lambda\, , \\
2\kappa^2\pi^S &=& \frac{\nabla^2B}{3a^2} + \ddot{B} + 3H\dot{B}\, ,\\
-\frac{\delta J_0}{\kappa^2} &=& \dot{\delta \rho} + 3H(\delta \rho + \delta p) + \nabla^2\left[ \frac{(\bar{\rho} + \bar{p})}{a^2}v + H\pi^S \right]\, ,\label{ecsg}\\
 \frac{\delta J^S}{\kappa^2} &=& \delta p + \nabla^2\pi^S + \partial_t\left[ (\bar{\rho} + \bar{p})v \right] + 3H(\bar{\rho} + \bar{p})v\, ,
\end{eqnarray}
\end{subequations}
with Eq.~\eqref{jmupert_syn} now written as
\begin{equation}
    \delta J_{\mu} = \frac{1}{4}\partial_{\mu}\left[ -\frac{2}{a^2}\nabla^2A + \kappa^2\left( 3\delta p - \delta\rho \right) \right]\, .
    \label{jmupert_syn_uni}
\end{equation}

As we have done in the previous case, it is possible to show that in a matter--dominated era the CDM density contrast in UG for the Synchronous gauge $\delta_{CDM(syn)}^{UG}$ is given by
\begin{equation}
     \delta_{cdm(syn)}^{{\rm{UG}}} = \frac{2k^2}{7a^2H^2}A\, ,
     \label{delta_cdm_ug_syn}
\end{equation}
where again we have used $\nabla \rightarrow -k^2$ for solutions in Fourier space. Then, once the solution for the gravitational potential $A$ is known, we have the cosmological evolution for the density contrast.

\subsection{Sachs--Wolfe effect: a proper derivation in UG}\label{Sachs--Wolfe effectug}

\qquad The approach followed by~\cite{Gao:2014nia,Basak:2015swx} lead to an expression that modifies the GR result by a new term (see Eq.~\eqref{sw}). However, we have shown in Section~\ref{alternative_gauge} that such new term is not completely determined unambiguously due to the gauge is not properly fixed. In what follows, we present the derivation for the Sachs--Wolfe effect for UG in the Newtonian gauge.

We start by setting the line element for the Newtonian gauge~\eqref{new_ds2}:
\begin{equation}
    ds^2 = -(1+2\Phi)dt^2 + a^2 (1-2\Psi)\delta_{ij}dx^idx^j\, ,
\end{equation}
where we left both gravitational potentials in order to compare with GR the final expression, and at the end of the procedure we use the unimodular constraint~\eqref{uc_ng}. Following~\cite{Dodelson:2003ft}, it can be shown that the Boltzmann equation for photons at linear order in UG is given by
\begin{equation}
    \frac{\partial }{\partial t}\left(\frac{\delta T}{\bar{T}}\right) + \frac{\hat{p}^{i}}{a}\frac{\partial }{\partial x^{i}}\left(\frac{\delta T}{\bar{T}}\right) - \frac{\partial \Psi}{\partial t} + \frac{\hat{p}^{i}}{a}\frac{\partial \Phi}{\partial x^{i}} = 0\, ,
\end{equation}
where the right hand side of the previous equation neglects the collision term since we are interested in the moment that photons are already decoupled. The mean temperature and its fluctuations are denoted respectively by $\bar{T}$ and $\delta T$, whereas $\hat{p}^{i}$ is the unitary 3--momentum. In order to apply the same differential operator to both gravitational potentials, we add new partial derivatives as follows
\begin{eqnarray}
    \frac{\partial }{\partial t}\left(\frac{\delta T}{\bar{T}}\right) + \frac{\hat{p}^{i}}{a}\frac{\partial }{\partial x^{i}}\left(\frac{\delta T}{\bar{T}}\right) - \frac{\partial \Psi}{\partial t} + \frac{\partial \Phi}{\partial t} + \frac{\hat{p}^{i}}{a}\frac{\partial \Phi}{\partial x^{i}} &=&  \frac{\partial \Phi}{\partial t}\nonumber \\
    \left( \frac{\partial }{\partial t} + \frac{\hat{p}^{i}}{a}\frac{\partial }{\partial x^{i}} \right)\left( \frac{\delta T}{\bar{T}} + \Phi\right) &=& \frac{\partial \Phi}{\partial t} + \frac{\partial \Psi}{\partial t} \, .
\end{eqnarray}

At this point, notice that once the gravitational potentials are equal the standard result is obtained, and the right hand side of the previous equation is $2\partial \Phi/\partial t$ (see Eq.(9.20) of~\cite{Mukhanov:2005sc}). Notwithstanding, the latter is true in GR where no anisotropic stress is present and the condition $\Phi = \Psi$ is satisfied. In our case, the anisotropic stress can not be set to zero in the Newtonian gauge, as we have discussed in previous Sections. Nonetheless, we have to impose the unimodular constraint of Eq.~\eqref{uc_ng} in the Newtonian gauge, which reads $\Psi = \Phi/3$. Thus, the relation between the temperature fluctuations $\delta T/\bar{T}$ and the gravitational potential $\Phi$ in UG is
\begin{equation}
    \left( \frac{\partial }{\partial t} + \frac{\hat{p}^{i}}{a}\frac{\partial }{\partial x^{i}} \right)\left( \frac{\delta T}{\bar{T}} + \Phi\right) = \frac{4}{3}\frac{\partial \Phi}{\partial t}\, ,
\end{equation}
and it is only a factor of $2/3$ the difference with respect to the GR result. After recombination the universe is matter--dominated and then we can approximate $\partial \Phi/\partial t \simeq 0$. This leads to the standard expression of the Sachs--Wolfe effect:
\begin{equation}\label{Sachs--Wolfe effect_ug}
    \left( \frac{\delta T}{\bar{T}} + \Phi\right) = {\rm{const}}\, .
\end{equation}

Therefore, whereas previous literature have found modifications due to the presence of a new gravitational potential in Eq.~\eqref{Sachs--Wolfe effect_ug} (see Eq.~\eqref{sw}), we have shown that such modification propagates spurious degrees of freedom due to the gauge choice. Our result shows that effectively there is no distinction between GR and UG when looking at the Sachs--Wolfe effect, but UG does not induced new terms in the relation between temperature fluctuations $\delta T/\bar{T}$ and the gravitational potential $\Phi$.

\section{Final remarks}\label{final_remarks}

\qquad The theory of Unimodular Gravity in its original formulation brings new interesting features due to the constraint in the spacetime four--volume, reducing general coordinate transformations to volume--preserving diffeomorphisms. The natural arising of the non--conservation of energy--momentum tensor allows to generate new non--gravitational interactions, which can be used to elucidate the behavior of the dark sector in cosmological models. 

We have analyzed whether the most common gauges used in cosmology are properly fixed, since previous work on linear perturbations within the framework of UG have not discussed this crucial aspect in the study of cosmological perturbations. We have demonstrated that it is possible to fix both, Newtonian and Synchronous gauges in UG, although the consequences on the matter fields are different to those for GR: particularly, CDM must have a non--null anisotropic stress when working in the Newtonian gauge, whereas it is not possible to choose a comoving observer with the CDM fluid in the Synchronous one.

Even when the dynamics of the perturbations change with respect to GR due to the unimodular constraint (we are left with only one gravitational potential instead of two), we have shown that it is possible to obtain the fluctuations of the CDM energy density as function of the only gravitational degree of freedom in both, Newtonian and Synchronous gauge. In fact, in the same line of ideas developed by Ma \& Bertschinger~\cite{Ma:1994dv}, we can obtain the equations in terms of the fluid variables: density contrast $\delta$ and velocity divergence $\theta$, given by
\begin{equation}
    \delta \equiv \frac{\delta \rho}{\bar{\rho}}\, , \quad \theta \equiv \frac{\partial_i v_i}{a} = \frac{1}{a}\partial_i (\partial_iv + v_i^V) = \frac{\nabla^2v}{a}\, ,
\end{equation}
where in the last expression, we only consider the scalar mode of peculiar velocity. From Eqs.~\eqref{emtng} and~\eqref{emtsg}, where the unimodular constraint has not been imposed yet, $\delta$ and $\theta$ are given in both gauges respectively by,

\textit{Newtonian gauge}:

\begin{subequations}\label{ng_dc_dv}
\begin{eqnarray}
     \delta^{\prime} &=& -(1+\omega)\left( \theta -3\Psi^{\prime} \right) - 3\frac{a^{\prime}}{a}\left( \frac{\delta p}{\delta \rho} - \omega \right)\delta + a\frac{\bar{J}_0 \delta - \bar{\rho}\delta J_0}{\kappa^2 \bar{\rho}}\, ,\\
     \theta^{\prime} &=& -\frac{a^{\prime}}{a}(1-3\omega)\theta - \frac{\omega^{\prime}}{1+\omega}\theta + \frac{\delta p/\delta \rho}{1+\omega}k^2\delta -k^2\sigma + k^2\Phi +\frac{a\bar{J}_0}{\kappa^2\bar{\rho}}\theta - \frac{k^2\delta J^S}{\kappa^2\bar{\rho}(1+\omega)}\, .
\end{eqnarray}
\end{subequations}

\textit{Synchronous gauge}:

\begin{subequations}\label{sg_dc_dv}
\begin{eqnarray}
     \delta^{\prime} &=& -(1+\omega)\left( \theta + \frac{h^{\prime}}{2} \right) - 3\frac{a^{\prime}}{a}\left( \frac{\delta p}{\delta \rho} - \omega \right)\delta + a\frac{\bar{J}_0 \delta - \delta J_0}{\kappa^2 \bar{\rho}}\, ,\\
     \theta^{\prime} &=& -\frac{a^{\prime}}{a}(1-3\omega)\theta - \frac{\omega^{\prime}}{1+\omega}\theta + \frac{\delta p/\delta \rho}{1+\omega}k^2\delta -k^2\sigma +\frac{a\bar{J}_0}{\kappa^2\bar{\rho}}\theta - \frac{k^2\delta J^S}{\kappa^2\bar{\rho}(1+\omega)}\, ,
\end{eqnarray}
\end{subequations}
where, for the sake of comparison with~\cite{Ma:1994dv}, this time the prime indicates derivative with respect to conformal time $\tau$, relating with cosmic time $t$ through $d\tau = dt/a$. We also identify our anisotropic term $\pi^S$ with $\sigma$ through the relation\footnote{In order to coincide in notation with~\cite{Ma:1994dv}, we have redefined the \textit{traceless} anisotropic stress by adding the term $-\delta_{ij}\nabla^2\pi^S/3$ in Eq.~\eqref{tijpertdc}.} $\sigma \equiv -2\nabla^2 \pi^S/3\bar{\rho}(1+\omega)$, and the trace part of the spatial metric perturbation in conformal time is related to our gravitational potentials in the synchronous gauge by $h = h_{ii} \equiv 3A + \nabla^2B$. Previous equations are the UG version of Eqs. (29) and (30) from~\cite{Ma:1994dv}, where new terms due to the energy--momentum current violation $J_{\mu}$ are present. From what we have learn in previous Sections, we have to set the unimodular constraint, and taking into consideration the new features arising in UG under the analysis of gauge fixing: for instance, in the Newtonian gauge we have to keep the anisotropic term in order to have gravity perturbations (see discussion in Section~\ref{lp_ng_sec}). On the other hand, once the synchronous gauge is fixed, it is not possible to have a comoving observer with the CDM fluid, and then, the velocity divergence can not be set equal to zero (see discussion in Section~\ref{lp_sg_sec}). Thus, considering the corresponding unimodular constraint in each gauge ($3\Psi = \Phi$ and $3A = -\nabla^2B$ for the Newtonian and Synchronous gauge respectively) for a CDM--dominated universe, the previous equations for the evolution of density contrast $\delta$ and velocity divergence $\theta$ are written as:

\textit{Newtonian gauge}:

\begin{subequations}\label{ng_dc_dv}
\begin{eqnarray}
     \delta^{\prime} &=& -\theta + \Phi^{\prime} + a\frac{\bar{J}_0 \delta - \bar{\rho}\delta J_0}{\kappa^2 \bar{\rho}}\, ,\\
     \theta^{\prime} &=& -\frac{a^{\prime}}{a}\theta - k^2\sigma + k^2\Phi +\frac{a\bar{J}_0}{\kappa^2\bar{\rho}}\theta - \frac{k^2\delta J^S}{\kappa^2\bar{\rho}(1+\omega)}\, .
\end{eqnarray}
\end{subequations}

\textit{Synchronous gauge}:

\begin{subequations}\label{sg_dc_dv}
\begin{eqnarray}
     \delta^{\prime} &=& -\theta + a\frac{\bar{J}_0 \delta - \delta J_0}{\kappa^2 \bar{\rho}}\, ,\\
     \theta^{\prime} &=& -\frac{a^{\prime}}{a}\theta +\frac{a\bar{J}_0\theta - k^2\delta J^S}{\kappa^2\bar{\rho}}\, .
\end{eqnarray}
\end{subequations}

In the case of Eqs.~\eqref{ng_dc_dv}, the differences with respect to GR and the standard $\Lambda$CDM model are the presence of the energy--momentum current violation $J_{\mu}$, the anisotropic term $\sigma$, and we have only one gravitational potential $\Phi$. Notice that even when assuming $\nabla_{\mu} T^{\mu \nu} = 0\, ,$ and then $\bar{J}_0 = \delta J_0 = \delta J^S = 0\, ,$ the perturbation equations do not recover the GR case. This is precisely due to both, the unimodular constraint and the anisotropic term. Similarly, the dynamics of the perturbed equations~\eqref{sg_dc_dv} for the Synchronous gauge in UG is very different from that of GR. Even when neglecting the energy--momentum current violation, we can observe that the source of the density contrast evolution is not the trace $h$ (as is the case in $\Lambda$CDM, see Eq.(42) in~\cite{Ma:1994dv}), but the velocity divergence $\theta$. This is another way to understand why it is not possible to choose an observer comoving with CDM: there will be not growth of structures if $\theta = 0$. Of course, in the general case we are studying, the presence of the energy--momentum current violation and its scalar perturbations will affect as well the dynamics of structure formation in both gauges. 

Therefore, UG leaves imprints in the properties of dark sector, and the implications on the linear perturbations are different to those in GR. Specifically, the physical consequences of linear perturbations in UG due to the geometric constraint imposed by the unimodular condition, are translated into non--standard features of the cold dark matter component: if working in the Newtonian gauge, CDM must be anisotropic as this term is directly proportional to the only gravitational degree of freedom (see Eqs.~\eqref{uc_ng} and~\eqref{phipsi_ug}); if working in the Synchronous gauge, it is not possible to choose a comoving observer with CDM fluid as its velocity divergence drives the structures formation (see Eqs.~\eqref{sg_dc_dv}). Besides, we have new contributions from the energy--momentum current violation, in the form of the background term $\bar{J}_0$ and scalar modes $\delta J_0\, , \delta J^S\, .$ The background dynamics have been solved and studied under numerical and statistical analysis, by considering phenomenological models of diffusion to describe the new non--gravitational interactions in the dark sector~\cite{LinaresCedeno:2020uxx}. However, the linear perturbations in UG lead to a novel level of complexity: we have focused in a matter--dominated universe in order to extract some information about the process of structure formation through the CDM density contrast, but this is not enough if one want to reproduce observables such as CMB or MPS. One of the new issues to handle is the information about the scalar perturbations of the energy--momentum current violation $\delta J_0$ and $\delta J^S$. Perhaps a naive way to proceed is to directly consider fluctuations of the diffusion models, or as was the case for the background in~\cite{LinaresCedeno:2020uxx}, to propose a phenomenological model for such perturbations. 

Thus, more work have to be done in order to properly implement a cosmological model based on UG in a Boltzmann solver such as \texttt{CAMB} or \texttt{CLASS}. Even when the conservative approach of energy--momentum conservation for ordinary matter content (photons, neutrinos, baryons) is assumed, the unimodular constraint changes the dynamics of linear perturbations for all species. In other words, while $J_{\mu} = 0$ for ordinary matter, the curvature produced by the only gravitational potential in UG will change the dynamics of matter fields. Even more, Boltzmann solvers are written for GR in the Synchronous gauge\footnote{See \url{https://cosmologist.info/notes/CAMB.pdf} and \url{http://www.class-code.net}. In particular, \texttt{CLASS} allows to work in both, Synchronous and Newtonian gauge. In any case, the numerical implementation must be applied consistently by considering the gravitational effects of the unimodular constraint at linear perturbations level for all matter components.}, and strictly speaking such gauge does not exist in UG, since it is not possible to choose consistently a CDM comoving observer by setting its velocity divergence $\theta_{CDM} = 0\, .$ With this in mind, we consider that for any attempt to reproduce CMB and MPS observations for cosmological models within the framework of UG, analysis of cosmological perturbations such as~\cite{Ma:1994dv,Bucher:1999re,Malik:2008im} must be done, in order to consistently solve the dynamics of linear perturbations for all matter and energy content gravitating as UG dictates. This work constitutes a first step in this direction by considering only the evolution of the dark matter component at linear order in perturbations.


\acknowledgments

F.X.L.C. acknowledges Beca CONACYT. U.N and F.X.L.C acknowledges to PROYECTO CIENCIA DE FRONTERA CF 2019/2558591 for financial support.  


\bibliographystyle{plunsrt}
\bibliography{bib}








\end{document}